\documentclass[aps,prb,showkeys,showpacs]{revtex4}
\usepackage{epsfig}
\usepackage{color}
\usepackage{graphicx}% Include figure files
\usepackage{dcolumn}% Align table columns on decimal point
\usepackage{bm}% bold math
\begin{document}
\pdfoutput=1
\title{Spatio-temporal behaviour of the deep chlorophyll maximum in Mediterranean Sea: Development of a stochastic model for picophytoplankton dynamics}
\author{\large G. Denaro$^a$, D. Valenti$^{a,1}$, A. La Cognata$^a$,
B. Spagnolo$^{a,2}$, A. Bonanno$^{b,3}$, G. Basilone$^b$, S.
Mazzola$^b$, S. Zgozi$^b$, S. Aronica$^b$, C.
Brunet$^c$\smallskip}\affiliation{{$^a$Dipartimento di Fisica, Universit\`a di Palermo}\\
{Group of Interdisciplinary Physics and CNISM, Unit\`a di Palermo}\\
{Viale delle Scienze, Ed.~18, I-90128~Palermo, Italy}\\
{$^1$e-mail: davide.valenti@unipa.it}\\
{$^2$e-mail: bernardo.spagnolo@unipa.it\smallskip}\\
{$^b$Istituto per l'Ambiente Marino Costiero, CNR, U.O.S. di Capo
Granitola, Via del Faro 3, I-91020 Campobello di
Mazara (TP), Italy.\\
{$^3$e-mail: angelo.bonanno@iamc.cnr.it\smallskip}\\
$^c$Stazione Zoologica Anton Dohrn, Villa Comunale, 80121 Napoli,
Italy.}}
\begin{abstract}
In this paper, by using a stochastic reaction-diffusion-taxis model,
we analyze the picophytoplankton dynamics in the basin of the
Mediterranean Sea, characterized by poorly mixed waters. The model
includes intraspecific competition of picophytoplankton for light
and nutrients. The multiplicative noise sources present in the model
account for random fluctuations of environmental variables.
Phytoplankton distributions obtained from the model show a good
agreement with experimental data sampled in two different sites of
the Sicily Channel. The results could be extended to analyze data
collected in different sites of the Mediterranean Sea and to devise
predictive models for phytoplankton dynamics in oligotrophic waters.
\end{abstract}
%
%\pacs{}

\keywords{Spatial ecology; Marine ecosystems; Phytoplankton
dynamics; Deep chlorophyll maximum; Random processes; Stochastic
differential equations}
\maketitle

\section{Introduction}
\label{intro} Natural systems are characterized by two factors: (i)
non-linear interactions among their parts; (ii) external
perturbations, both deterministic and random, coming from the
environment~\citep{Spa04,Hup05,Ebe05,Pro08,Spa08,Val08}. It is worth
noting that natural systems, because of these characteristics, are
complex systems~\citep{Grenfell,Zimmer,Bjornstad,Spa02,Lab02,Spg02,
Spa03,Spa05,Car05,Chi05,Fia06,Val06,Chichigina}. Therefore, the
study of a marine ecosystem has to be performed by considering the
perturbations, not only deterministic but also random, due to the
fluctuations of the environmental variables. This implies the
necessity of including in the model a term which describes the
continuous interaction between the ecosystem and environment. In
particular, physical variables, such as temperature, salinity and
velocity field, are affected by random perturbations and can be
therefore treated as noise sources. This causes the phytoplankton
behaviour to be subject to a stochastic dynamics, and allows to
expect that a stochastic approach should reproduce the distributions
of phytoplankton biomass better than deterministic models. On this
basis, noise effects have to be included to better analyze the
dynamics of a marine system such as
that studied in this work.\\
The growth of phytoplankton is limited by the concentration of
nutrients $R$ and intensity of light $I$~\citep{Kla01,Kla07}. In
particular, the survivance of phytoplankton is strictly connected
with the presence of sufficiently high nutrient concentration. It is
worth stressing that nutrients, which are in solution, diffuse from
the bottom (seabed) towards the top (water surface). Nutrient
distributions along the water column are therefore characterized by
an increasing trend from the sea surface to the benthic layer. As a
consequence, the positive gradient of nutrient concentration causes
the maxima of chlorophyll, which is contained in the phytoplankton
cells, to be localized in deep subsurface layers. This condition
constitutes one of the most striking feature of the nutrient poor
waters in ocean ecosystems and freshwater
lakes~\citep{And69,Cul82,Abb84,Tit03}. Conversely, the light
penetrates through the surface of the water and has an exponentially
decreasing trend along the water column. This characteristic makes
the deep layers unfavourable for the photosynthesis, determining, as
a consequence, adverse life conditions for phytoplankton. In
particular, light is a crucial parameter for the localization of the
deep chlorophyll maximum (DCM), as revealed by the significant
correlation found between the depth of DCM and light intensity over
the Mediterranean basin in summer (Brunet et al., unpublished
data).\\
The dynamics, competition and structuring of phytoplankton
populations have been investigated in a series of theoretical
studies based on model
systems~\citep{Rad75,Var92,Hui95,Kla01,Die02,Hod04,Bec07,Kla07,Mei09,Bou10}.
In a few recent investigations it was observed that in the presence
of an upper mixed layer either surface or deep maxima can be
observed indifferently under almost the same
conditions~\citep{Ven93,Hol04,Rya10}.\\
In view of analyzing an ecological system, as a preliminary step it
is necessary to define the correct values of the parameters and the
role that they play on the dynamics of the populations, specifically
when the coexistence of different species in the same community is
considered~\citep{Nor04}. The responses of the species to
environmental solicitations strongly depend on the biological and
physical parameters. Among these, a relevant role is played by the
phytoplankton velocity which is strictly connected with the
microorganism size, one of the main functional traits for
phytoplankton diversity. Other parameters that influence the balance
of a marine ecosystem are, for example, growth rates and
nutrient uptake~\citep{Fog91,Pre91}.\\
In this paper we deal with data obtained in a hydrologically stable
area of the Mediterranean Sea, where the environmental light and
nutrients, specifically phosphorus, contribute to determine life
conditions. The Mediterranean basin is characterized by oligotrophic
conditions and it has been suggested that there is a decreasing
trend over time in chlorophyll concentration. This has been
associated with increased nutrient limitation resulting from reduced
vertical mixing due to a more stable stratification of the basin, in
line with the general warming of the Mediterranean ~\citep{Bar08}.
Here we consider the Strait of Sicily, which is known to govern the
exchanges between the eastern and western basins and is
characterized by active mesoscale dynamics~\citep{Ler01}, strongly
influencing the ecology of phytoplankton communities. Moreover, the
Strait of Sicily is a biologically rich area of the Mediterranean
Sea with a key role in terms of fisheries~\citep{Gar02,Cut03}. The
anchovy growth (along with phytoplankton biomass) in the Sicilian
Channel resulted to be mainly explained by changes in the
chlorophyll concentration, used as a phytoplankton biomass
indicator~\citep{Bas04}. Our study is performed using a stochastic
model obtained by modifying a deterministic reaction-diffusion-taxis
model. Specifically, the analysis focuses on the spatio-temporal
dynamics of the phytoplankton biomass, and provides the time
evolution of biomass concentration along the water column. Finally,
the results are compared with experimental data collected in two
different sites of the Strait of Sicily.
\section{Materials and methods} \label{Mat_methods}
\subsection{Environmental data} \label{Env_data}
\begin{figure}[htbp]
\centering
\includegraphics[width=8cm]{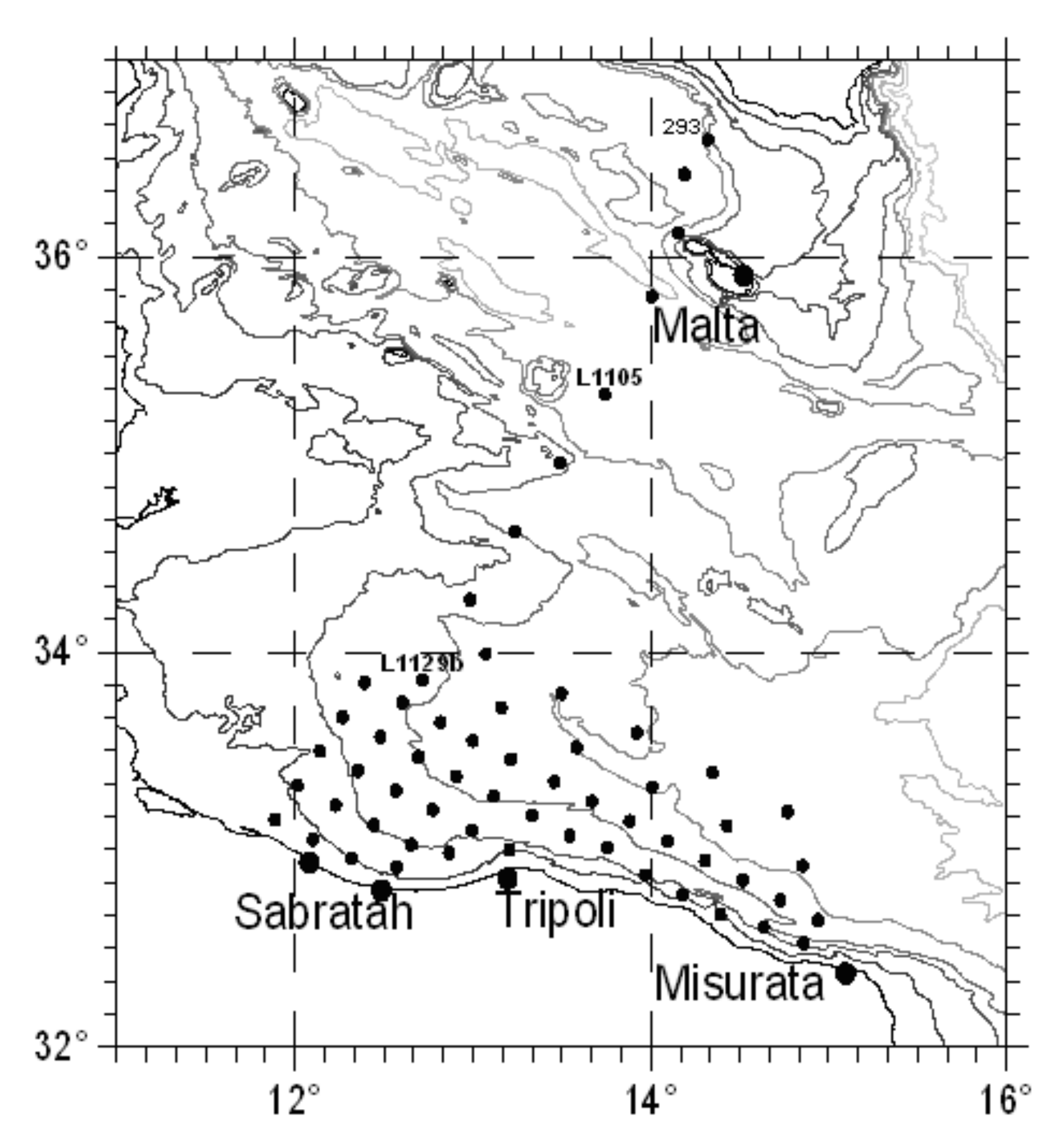}
\caption{Locations of the CTD stations where the experimental data
were collected.} \label{fig:location}
\end{figure}

The experimental data were collected in the period 12$^{th}$ -
24$^{th}$ August 2006 in the Sicily Channel area (Fig.
\ref{fig:location}) during the MedSudMed-06 Oceanographic Survey
onboard the R/V Urania. Hydrological data were obtained using a
SBE911 plus CTD probe (Sea-Bird Inc.); chlorophyll a fluorescence
data (\emph{chl a}, $\mu$g/l) were contemporary acquired by means of
the Chelsea Aqua 3 sensor. In the Libyan area the CTD stations were
located on a grid of 12 x 12 nautical miles. Moreover, CTD data have
been collected along a transect between the Sicilian and the Libyan
coasts. In the present work, two stations out of the whole data set
were considered. The selected stations were located on the south of
Malta (site L1105) and on the Libyan continental shelf (site
L1129b).
The collected data were quality-checked and processed following the
MODB instructions~\citep{Bra94} using Seasoft software. The
post-processing procedure generated a text file for each station
where the values of the oceanographic parameters were estimated with
a 1 m step. Hydrological conditions remained constant for the entire
sampling period and were representative of the oligotrophic
Mediterranean Sea in summer. Nitrate, nitrite, silicate and
phosphate concentrations were not determined.
\subsection{Phytoplanktonic data} \label{Phyto_data}
\begin{figure}[htbp]
\centering
\includegraphics[width=8cm]{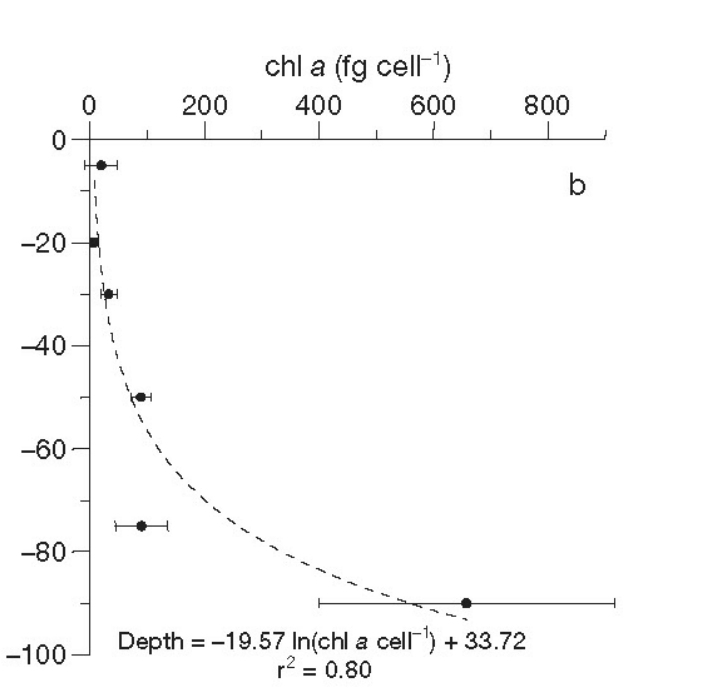}
\caption{Mean vertical profile of \textit{chl a} per picoeukaryote
cell (fg cell$^{-1}$). Error bars are Standard Deviation. Equation
and $r^2$ for the fit are reported on the plots. (Courtesy of Brunet
et al., 2007).}\label{fig:espon} \vskip-0.3cm
\end{figure}

Depending on size the phytoplankton species can be divide into two
main fraction:
\begin{itemize}
\item $<3 \mu m$ picophytoplankton, formed by groups,
Prochlorococcus, Synechococcus and
picoeukaryotes~\citep{Ols93,Bru08}. This size of phytoplankton
accounts for about 80\% of the total \textit{chl a} on average
~\citep{Bru06}, ranging from 40\% to 90\% (69\% in the
DCM)~\citep{Bru07}.
\item $>3 \mu m$  nano- and micro-phytoplankton, characterized by a
lower correlation with nutrients and salinity respect to
picophytoplankton. This is connected with the fact that the
contribution of picophytoplankton in the DCM is higher than in the
surface layer~\citep{Bru06}. This larger size fraction of
phytoplankton amounts to 20\% of the total \textit{chl a} on average
and is uniformly distributed along the water column.
\end{itemize}
The high pigment diversity of the smaller phytoplankton in the DCM
and its elevated contribution to the total \textit{chl a} indicated
a strong degree of adaptation to the quantity and quality of light
available~\citep{Dim07,Bru08,Dim08}. This is not true for the larger
phytoplankton, which is represented mainly by diatoms or
Haptophytes. Picoeukaryotes, which belong to the smaller size class,
present peculiar eco-physiological
properties~\citep{Rav05,Dim07,Wor08}, such as low sinking, high
growth rate and low nutrient uptake. Their small size leads to a low
package effect, which contributes to the light-saturated rate of
photosynthesis that can be achieved at relatively low
irradiances~\citep{Rav98,Bru03,Rav05,Fin05}. Due to their
peculiarities and relevant role in ecosystem functioning, they
constitute  a key-group to be considered within a model of
population dynamics. In Sicily Channel~\citep{Cas03,Bru06,Bru07},
picophytoplankton is numerically dominated by the Prochlorococcus
fraction. In this area the number of Prochlorococcus cells is
constant in the first $20$ m, and is characterized in the DCM by an
average value of $5.2 \times 10^4$ cell ml$^{-1}$. Average
picoeukaryote concentration in the DCM is $0.6 \pm 0.4 \times 10^3$
cell ml$^{-1}$, and the mean value of \textit{chl a} cell$^{-1}$
ranges between 10 and 660 fg \textit{chl a} cell$^{-1}$ along the
water column, with a significant exponential increase with depth
(see Fig.~\ref{fig:espon})~\citep{Bru07}. \vspace{0.0cm} The
concentration of \emph{chl~a} (fg cell$^{-1}$) per cell in
picoeukaryotes was highly variable among different water masses,
with significantly higher values in the DCM respect to the surface,
as a result of photoacclimation to decreased light
irradiances~\citep{Bru03,Dim07,Bru08,Dim08}.
\section{Experimental results} \label{Results}
\begin{figure}[htbp]
\centering \vspace{-0.0cm}
\includegraphics[width=10cm]{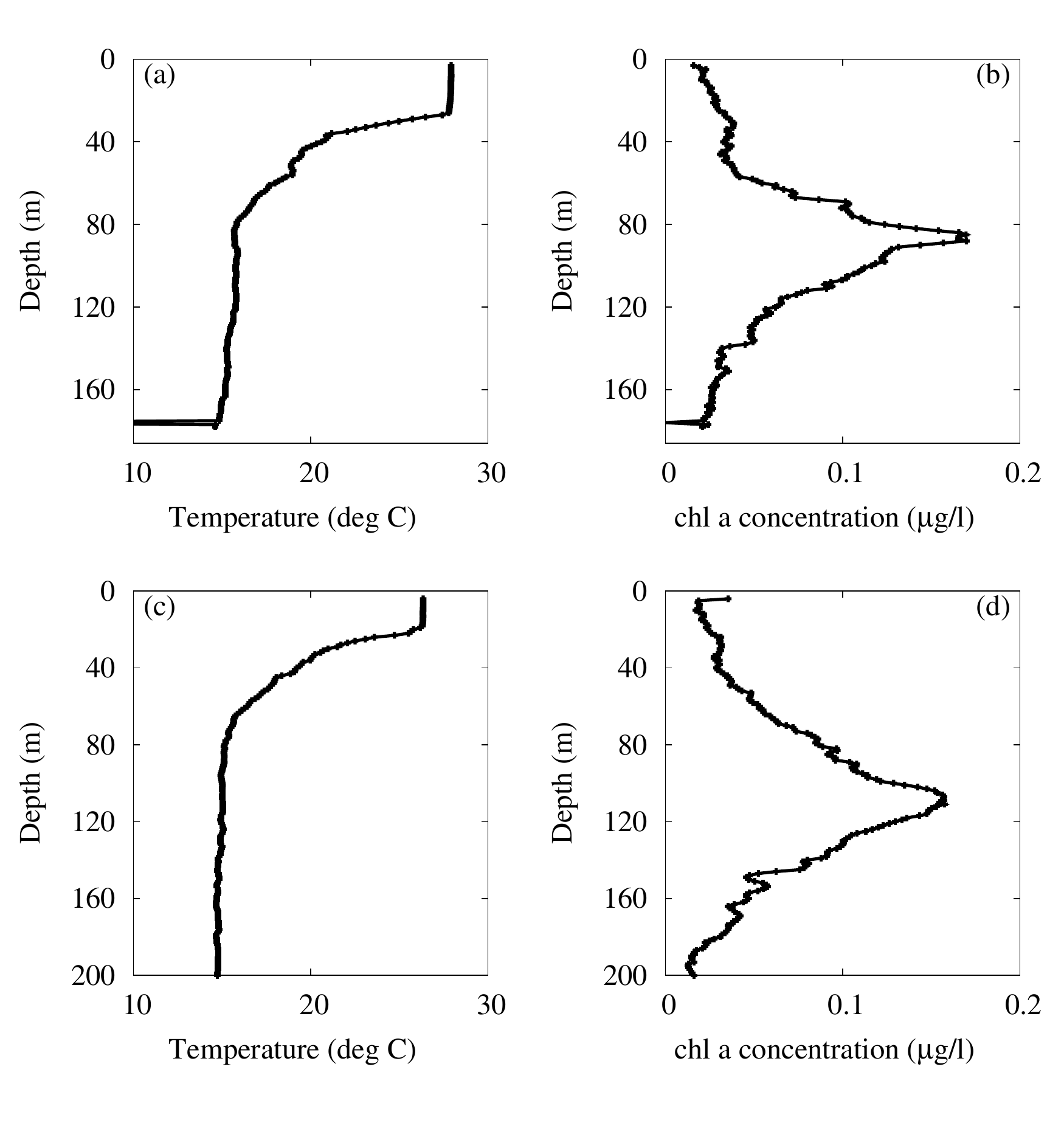}
\caption{Profiles of temperature (panels a, c) and \emph{chl a}
concentration (panels b, d) measured in sites L1129b and L1105. The
black lines have been obtained by connecting the experimental points
corresponding to samples distanced of 1 meter along the water
column. The total number of samples measured in the two sites is $n
= 176$ for L1129b, and $n = 563$ for L1105.}\label{fig:data_staz1}
\end{figure}

Data obtained from the cruises in two different sites of the Strait
of Sicily both for temperature and \emph{chl a} concentration are
shown in Fig.~\ref{fig:data_staz1}.
In site L1129b, the behaviour of the temperature along the water
column indicates the presence of a mixed layer (from the surface to
28 m depth) characterized by a high value of temperature. Below the
thermocline (28 m depth) the temperature decreases up to 80 m,
becoming uniform below this depth (Fig.~\ref{fig:data_staz1}a). The
site L1105 shows a mixed layer over the first 24 m of depth, and a
sharp decrease of temperature from 24 to 75 m
(Fig.~\ref{fig:data_staz1}c). Experimental data for \emph{chl a}
concentration show a nonmonotonic behaviour, as a function of the
depth, characterized by the presence of DCM in both sites (see
Fig.~\ref{fig:data_staz1}b,d). Specifically, fluorescence profiles
show a similar behaviour in the two sites, with \emph{chl a}
concentration ranging between $0.010$ and $0.17 \mu g$ \textit{chl
a} \textrm{l}$^{-1}$. Differences between the two sites are observed
in the depth, shape and width of the DCM.
\section{The Model}
\label{model}
%%%Ryabov 2010
In this study we analyze the spatio-temporal dynamics of a
picophytoplankton community, limited by nutrient and light in a
vertical poorly mixed water column. The mechanism, responsible for
the phytoplankton dynamics, is schematically shown in Fig. 4.
\begin{figure}[htbp]
\centering
\includegraphics[width=9.5cm,height=7cm]{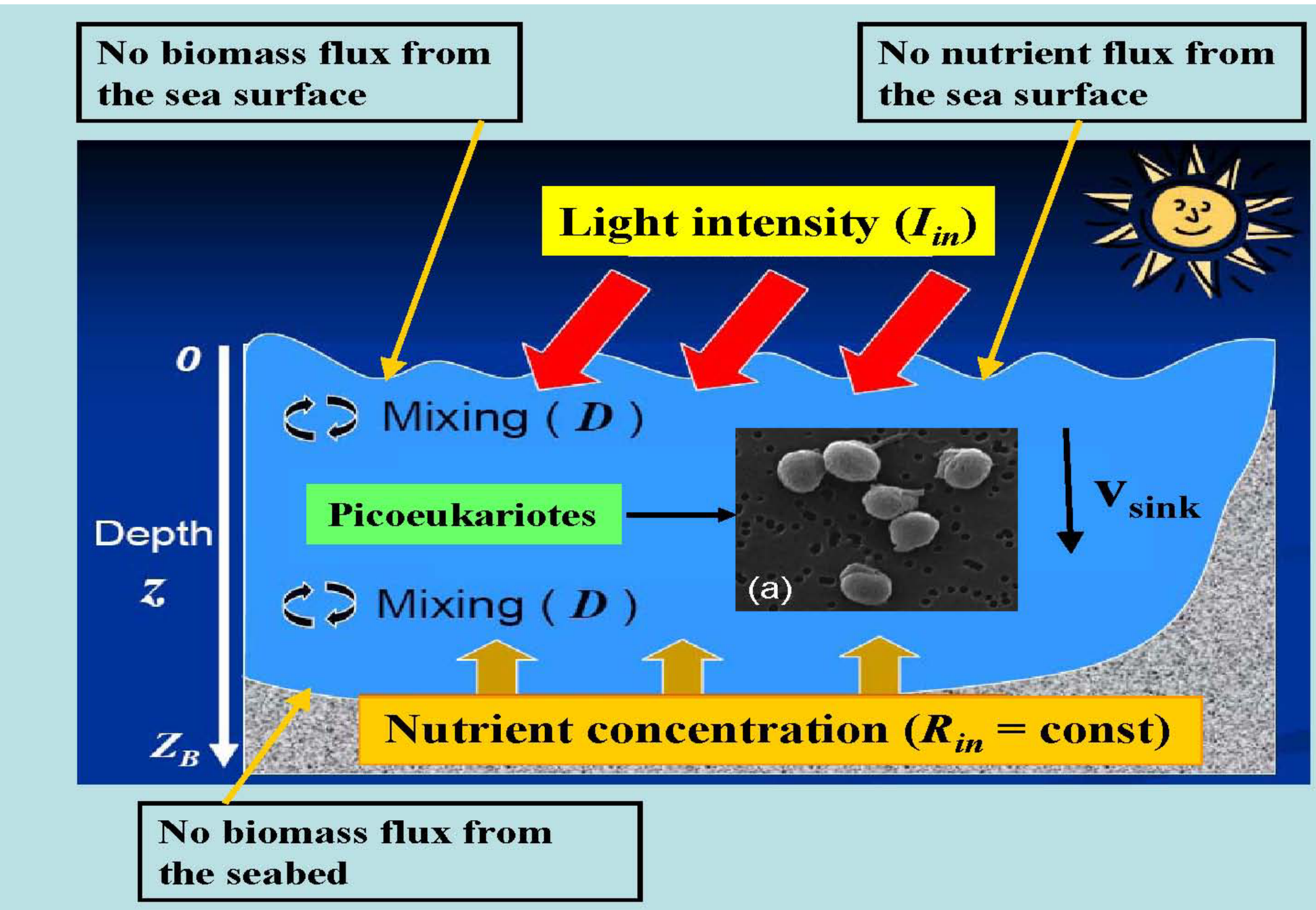}
\caption{(Color online) Scheme of the mechanism responsible for the
phytoplankton dynamics (modified from original figure by A. Ryabov).
(a) Image of Micromonas NOUM17 (courtesy of Augustin Engman, Rory
Welsh, and Alexandra Worden.}\label{scheme}
\end{figure}
The mathematical tool used to simulate the phytoplankton dynamics is
an advection-reaction-diffusion model. In particular, we investigate
the distribution of the picophytoplankton along the water column,
with light intensity decreasing and nutrient concentration
increasing with depth. Analysis and numerical elaborations are
divided in two phases:
\begin{itemize}
\item \textit{Phase 1}. By using a model based on two differential equations,
the distribution of picophytoplankton biomass $b$ is obtained along
the poorly mixed water column as a function of the time and depth,
and simultaneously the distribution of nutrient concentration $R$,
which limits the growth of phytoplankton, is calculated. The results
obtained are compared with the experimental data collected in the
two different sites of the Strait of Sicily.
\item \textit{Phase 2}. In order to match better the results for $b$
and $R$ to the experimental data, the random fluctuations of the
environmental variables are taken into account. In particular, a
stochastic model is obtained from the deterministic one by inserting
into the equations terms of multiplicative Gaussian noise.
\end{itemize}
\subsection{The deterministic model}
\label{det_mod} \label{det_mod} Here we introduce the model
consisting of a system of differential equations, with partial
derivatives in time and space (depth). The model allows to obtain
the dynamics of the phytoplankton biomass $b(z,t)$ and nutrient
concentration $R(z,t)$. The light intensity $I(z,t)$ is given by a
function varying, along the water column, with the depth and biomass
concentration. The behaviour of the phytoplankton biomass, along the
water column, is the results of three processes: growth, loss, and
movement. The phytoplankton growth rate depends on $I$ and
$R$~\citep{Kla01,Kla07,Mei09,Bou10,Rya10}. The limitation in
phytoplankton growth is described by the Monod
kinetics~\citep{Tur88}. The gross phytoplankton growth rate per
capita is given by $\min\{f_{I}(I),f_{R}(R)\}$, where $f_{I}(I)$ and
$f_{R}(R)$ are obtained by the Michaelis-Menten formulas
\begin{eqnarray}
&&f_{I}(I)=rI/(I+K_{I})\label{michelis1},\\
&&f_{R}(R)=rR/(R+K_{R})\label{michelis2}.
\end{eqnarray}
In Eqs.~(1),~(2),  $r$ is the maximum growth rate, while $K_I$ and
$K_R$ are the half-saturation constants for light intensity and
nutrient concentration, respectively. Varying $K_R$ and $K_I$ allows
to model, for instance, a species which is better adapted to the
light (smaller values of $K_I$) or nutrient (smaller values of
$K_R$). More specifically, we consider a species with small $K_I$
and large $K_R$ that corresponds to good life conditions at large
depth. These constants depend on the metabolism of the specific
microorganism considered.\\
The biomass loss, connected with respiration, death, and grazing,
occurs at a rate $m$~\citep{Kla01,Hui06,Rya10}. The
gross per capita growth rate is defined as\\
\begin{equation}
g(z,t)=\min(f_{R}(R(z,t)),f_{I}(I(z,t))). \label{g-rate}
\end{equation}
Turbulence, responsible for passive movement of the phytoplankton,
is modeled by eddy diffusion. Specifically, we describe turbulence
assuming that the vertical diffusion coefficient is uniform with the
depth and characterized by a low value ($D_b=D_R=0.5$). This choice
is motivated by the fact that in sites L1129b and L1105 the
phytoplankton peaks, located at 87 m and 111 m respectively, are
quite far from the thermocline (see Fig.~\ref{fig:data_staz1}).
Therefore, phytoplankton should go up (or down) if the biological
conditions are more suitable for growth above (below) than below
(above). Finally, no migration should occur if the biomass
concentrations are the same at different depths. These assumptions
about growth, loss, and movement, allow to obtain the following
differential equation for the dynamics of biomass concentration
$b$~\citep{Kla01,Hui06}:\\
\begin{equation}
\frac{\partial b(z,t)}{\partial
t}=g(z,t)b(z,t)-mb(z,t)+D_{b}\frac{\partial^{2} b(z,t)}{\partial
z^{2}}-v\frac{\partial b(z,t)}{\partial z}. \label{evoluzb}
\end{equation}
The positive phytoplankton velocity \emph{v}, due to active
movement, is oriented downward (sinking), in the direction of
positive \emph{z}. Phytoplankton does not enter or leave the water
column. This is set by using no-flux boundary conditions at $z=0$
and $z=z_{b}$:
\begin{equation}
\left.\left[D_{b}\frac{\partial b}{\partial
z}-vb\right]\right|_{z=0}= \left.\left[D_{b}\frac{\partial
b}{\partial z}-vb\right]\right|_{z=z_{b}}=0. \label{contornob}
\end{equation}
Eddy diffusion is responsible for mixing of the nutrient
concentration along the water column, with diffusion coefficient
$D_{R}$. The nutrient consumed by the phytoplankton is also obtained
from recycled dead phytoplanktonic microorganisms. The dynamics of
nutrient concentration can be therefore modeled as follows:
\begin{equation}
\frac{\partial R(z,t)}{\partial
t}=-\frac{b(z,t)}{Y}g(z,t)+D_{R}\frac{\partial^{2} R(z,t)}{\partial
z^{2}}+\varepsilon m\frac{b(z,t)}{Y}, \label{evoluzR}
\end{equation}
Here $Y$ is the phytoplankton produced biomass per unit of consumed
nutrient, and $\varepsilon$ is the nutrient recycle coefficient.
Since the nutrient is not supplied by the sea surface but comes from
the seabed, its concentration is set to the constant value $R_{in}$
in the sediment and, as a consequence, to the value $R(z_{b})$ in
the bottom of the water column. In fact the nutrient diffuses across
the sediment-water interface with a rate proportional to the
concentration difference between the solid phase (seabed) and the
deepest water layer (bottom of the water column).
Accordingly, the boundary conditions are given by:\\
\begin{equation}
\left.\frac{\partial R}{\partial
z}\right|_{z=0}=0,\quad\left.\frac{\partial R}{\partial
z}\right|_{z=z_{b}}=h(R_{in}-R(z_{b})), \label{contornoR}
\end{equation}
where $h$ is the permeability of the interface. Finally, taking into
account Lamber-Beer's law~\citep{Shi81,Kir94}, the light intensity
is characterized by an exponential decrease modeled as follows:
\begin{equation}
I(z)=I_{in}\exp\left\{-\int_0^z \left[ab(Z)+a_{bg}\right]
dZ\right\}, \label{evoluzI}
\end{equation}
where $a$ and and $a_{bg}$ are phytoplankton biomass and background
attenuation coefficients, respectively. Equations
(\ref{evoluzb})-(\ref{evoluzI}) form the biophysical model used in
our study.
\subsection{Results of the deterministic model}
\label{res_det_mod} 
\begin{figure}[htbp]
\centering
\includegraphics[width=10cm]{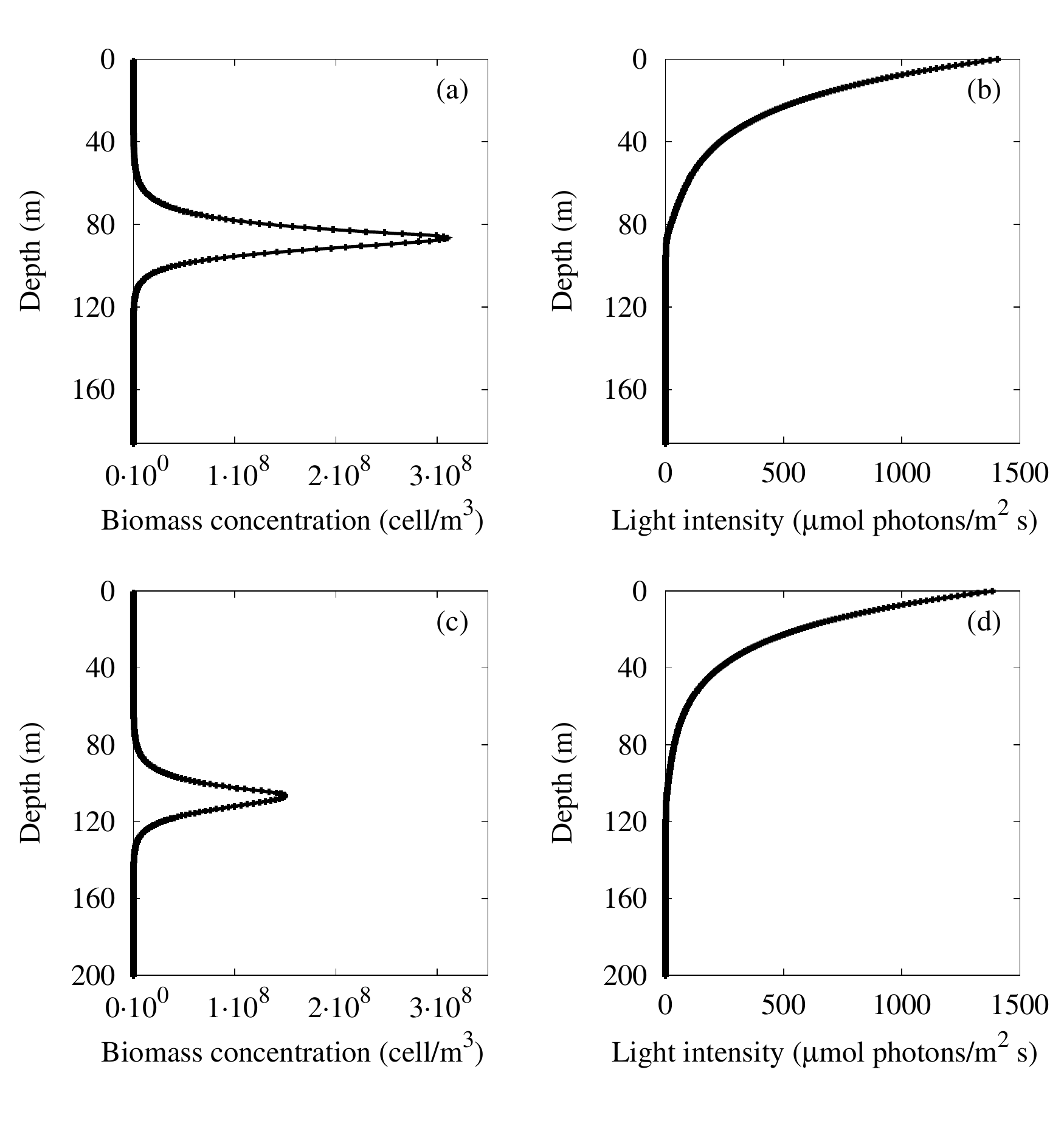}
\caption{Stationary distributions of the biomass concentration and
light intensity in sites L1129b (panels a, b) and L1105 (panels c,
d) as a function of depth.}\label{fig:confrontobI}
\end{figure}

The time evolution of the system is studied by
analyzing the spatio-temporal dynamics of biomass and nutrient
concentrations. In particular, by using a numerical method,
implemented by a program in C++ language and based on an explicit
finite difference scheme, equations (\ref{evoluzb})-(\ref{evoluzI})
are solved. The increment of the spatial variable is set to 0.5 m.
In view of reproducing the spatial distributions observed in the
real data for the phytoplankton biomass (see
Fig.~\ref{fig:data_staz1}), we choose the values of the
environmental and biological parameters to satisfy the monostability
condition corresponding to the presence of a deep chlorophyll
maximum~\citep{Kla01,Hui06,Rya10}. The numerical values assigned to
the parameters are shown in Table~\ref{table1}.
\begin{table*}[htbp]
\centering
\includegraphics[width=14cm]{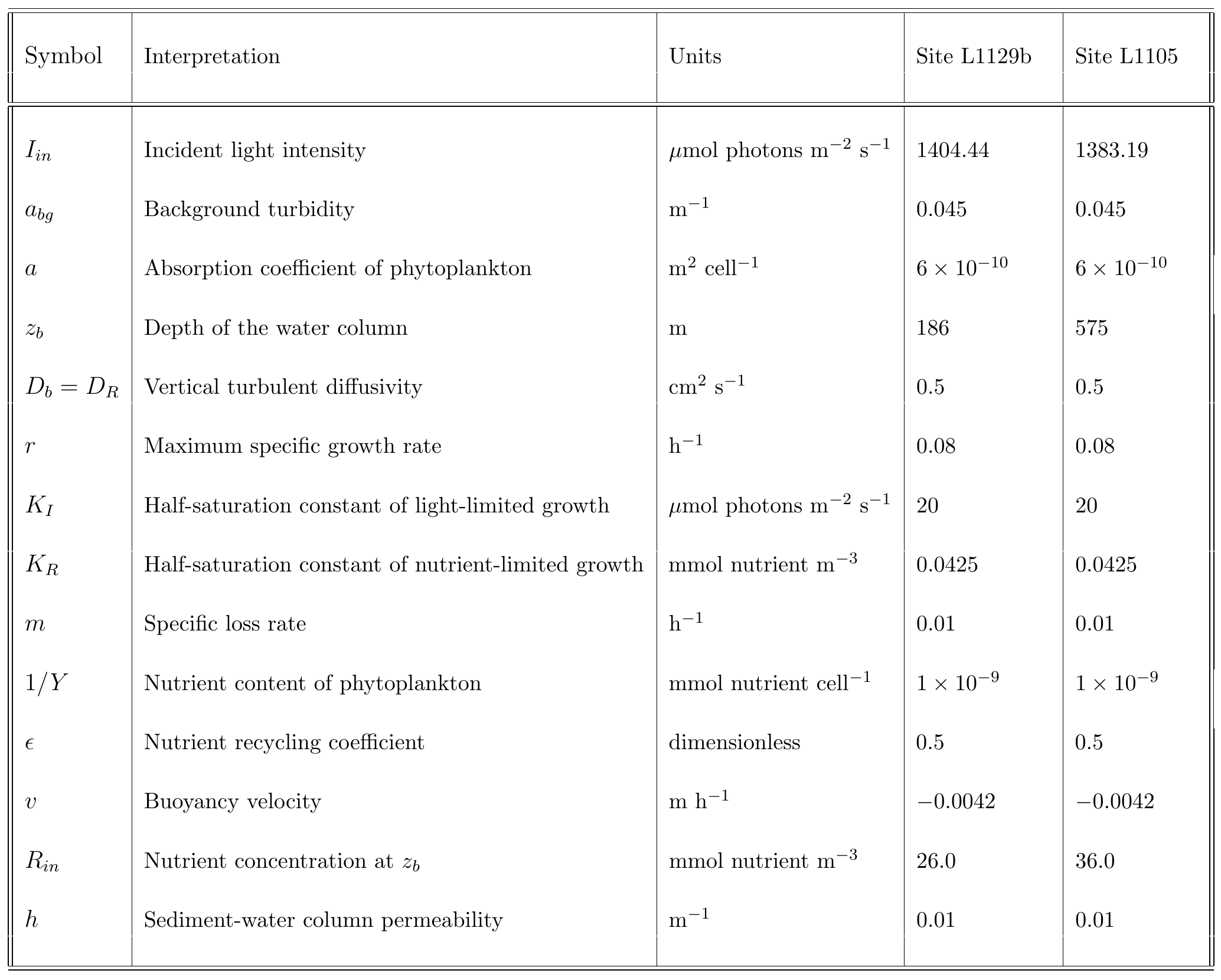}
\caption{Parameters used in the model. The values of the biological
and environmental parameters are those typical of picophytoplankton
and summer period in Mediterranean Sea, respectively.}\label{table1}
\end{table*}
Specifically, the values of the biological parameters $r$, $K_I$,
$K_R$, $v$, have been chosen to reproduce the behaviour of
picoeukaryotes. We note that, in systems characterized by a constant
value of the diffusion coefficient, the stationary state does not
depend on the initial conditions, according to previous
studies~\citep{Kla01,Rya10}. In order to obtain the steady spatial
distribution, we integrated numerically our equations over a time
interval long enough to observe the stationary solution. As initial
conditions we consider that the phytoplankton biomass is
concentrated in the layer where the maximum of the experimental
chlorophyll distribution is observed. On the other side the nutrient
concentration is approximately constant from the water
surface to the DCM, and increases linearly below this point up to the seabed.\\
Preliminary analysis (data not shown) revealed that the stationary
solution is characterized by DCMs which are shallower as the
nutrient supply increases, and deeper for enhanced light radiation.
In general, large values of $I_{in}$ (incident light intensity at
the water surface) lead to stationary conditions characterized by
DCM, while large values of $R_{in}$ (nutrient concentration in the
sediment) determine an upper chlorophyll maximum (UCM). Finally, for
intermediate values of $I_{in}$ and $R_{in}$ the chlorophyll maximum
can be localized close to the surface or at different depths,
depending on the values of the other parameters~\citep{Rya10}.\\
\indent In our study the values of the light intensity resulted to
be quite high in both sites, since sampling occured during summer
(August 2006). In this period the light intensity at the water
surface is larger than $1300$ $\mu$mol photons m$^{-2}$ s$^{-1}$.
Moreover the sinking velocity is set to the value typical for
picophytoplankton, $v=0.1$ m day$^{-1}$~\citep{Hui06}. The diffusion
coefficent is fixed at the value $D_b=0.5$ cm$^2$ sec$^{-1}$, which
corresponds to the condition of poorly mixed waters. By solving
Eqs.~(\ref{evoluzb})-(\ref{evoluzI}) we obtain the biomass
concentration expressed in cells/m$^3$ along the water column.
Depths of the water column used in the model were set according to
the measured depths in the corresponding marine sites. Moreover the
light intensities, $I_{in}$, are fixed using data available on the
NASA web site\footnote{http://eosweb.larc.nasa.gov/sse/RETScreen/}.
Finally, nutrient concentrations at the seabed were set at values
such as to obtain, for each site, a peak of biomass concentration at
the same position of the peak experimentally observed. All the other
parameters are the same in both sites. The growth rate obtained from
Eq.~(\ref{g-rate}) agrees with the values measured by other authors~\citep{Dim09}.\\
We note that our numerical results were obtained using a maximum
simulation time $t_{max} = 10^{5}~h$. Simulations (here not
reported) performed within the deterministic approach show that the
stationary regime is reached at $t\approx3\cdot 10^4~h$. This
indicates that, to reach the steady state, it is sufficient to solve
the equations of our model with a maximum time $t_{max} = 4\cdot
10^4~h$. By this way, we get the stationary profiles, both for
biomass concentration and light intensity, shown in
Fig.~\ref{fig:confrontobI}. Here we can note the presence of a
biomass peak as found in the experimental data, and the typical
exponential behaviour of the light intensity.
\begin{figure}[htbp]
\centering
\includegraphics[width=10cm]{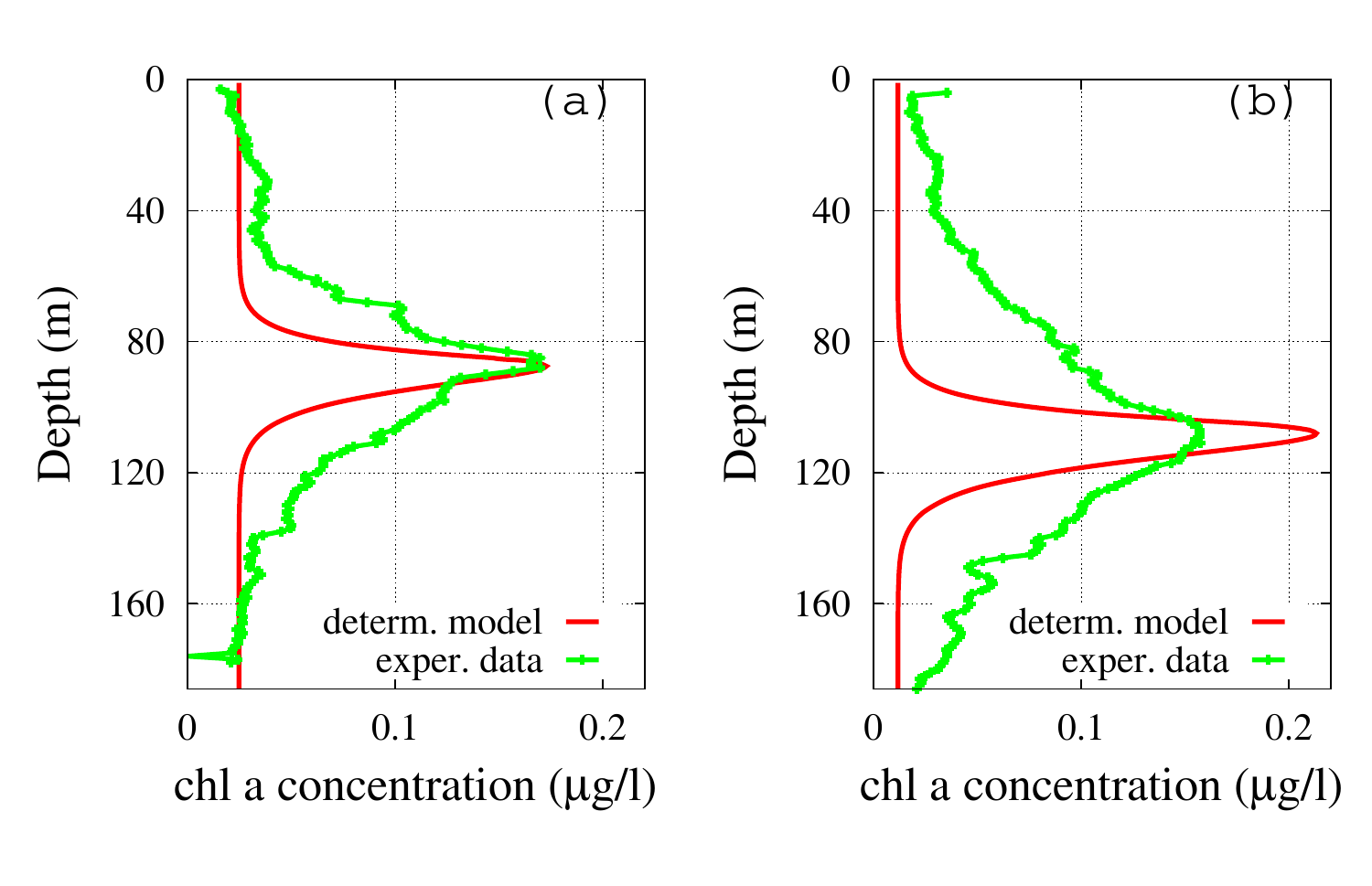}
\caption{(Color online). Stationary distribution of the chlorophyll
a concentration as a function of depth calculated (red line) by the
deterministic model and measured (green line) in sites (a) L1129b
and (b) L1105.}\label{fig:confronto1}
\end{figure}
To compare the theoretical results with the experimental data, we
exploit the curve of Fig.~\ref{fig:espon} to convert the cell
concentrations, obtained from the model and expressed in cell/m$^3$,
into \emph{chl a} concentrations expressed in $\mu$g/l. We recall
that about 43\% of the total quantity of \emph{chl
a}~\citep{Hui06,Bru06} is due to nano- and micro-phytoplankton (20\%
of the total \textit{chl a} on average), and Synechococcus (23\% of
the total \textit{chl a} on average), quite uniformly distributed
along the water column. Since our model accounts for the dynamics of
picoeukaryotes, to compare the numerical results with the
experimental data, we consider the 43\% of the total biomass and
divide it by depth, obtaining for each site the value $\Delta
b_{chl\thinspace a}$, which represents a constant concentration due
to other phytoplankton species present in the water column. Finally
along the water column we add the theoretical concentration with
$\Delta b_{chl\thinspace a}$ and obtain, for the distributions of
\textit{chl a} concentration, the stationary theoretical profiles
consistent with those of the experimental data. The results are
shown in Fig.~\ref{fig:confronto1}. Here we can observe that in both
sites the deep chlorophyll maxima obtained from the model are
located at the same depth of those observed experimentally. However,
the shape of the theoretical \emph{chl a} distributions is quite
different from the experimental profiles. Finally, we note that in
site L1105 the magnitude of the theoretical DCM is significantly
different from that observed in real data.

\subsection{The Stochastic Model}
In the previous section we used a deterministic model to fit the
experimental distributions of \emph{chl a} concentration. The
results obtained reproduce partially the characteristics of the
experimental profiles. In order to get a good agreement between real
data and theoretical results, we recall that the sea is a complex
system. This implies, as discussed in Introduction, the presence of
non-linear interactions among its
parts~\citep{Spa04,Hup05,Ebe05,Pro08,Spa08,Val08} and a continuous
interaction between the ecosystem and environment. In particular,
the system dynamics is affected not only by deterministic forces but
also random perturbations coming from the environment. In this
context environmental variables, due to their random fluctuations,
can act as noise sources, causing phytoplankton to be subject to a
stochastic dynamics. Therefore, in order to perform an analysis that
takes account for real conditions of the ecosystem, it is necessary
to modify our model, including the noise effects.
In the following we analyze two different situations.\\
\textbf{Case 1.} The environmental noise affects only the biomass
concentration. Therefore, Eqs.~(\ref{contornob})-(\ref{evoluzI}) are
maintained unaltered, while Eq.~(\ref{evoluzb}) becomes\\
\begin{equation}\label{evoluzbrum}
\frac{\partial b}{\partial t}=gb-mb+D_{b}\frac{\partial^{2}
b}{\partial z^{2}}-v\frac{\partial b}{\partial z}+b\thinspace
\xi_b(z,t)
\end{equation}
\textbf{Case 2.} The environmental noise affects only the nutrient
concentration. In this case,
Eqs.~(\ref{evoluzb}),(\ref{contornob}),(\ref{contornoR}),(\ref{evoluzI})
are maintained unaltered, while Eq.~(\ref{evoluzR}) is replaced by\\
\begin{equation}\label{evoluzRrum}
\frac{\partial R}{\partial
t}=[m\varepsilon-g]\frac{b}{Y}+D_{R}\frac{\partial^{2} R}{\partial
z^{2}}+R\thinspace \xi_R(z,t).
\end{equation}
In Eqs.~(\ref{evoluzbrum}) and (\ref{evoluzRrum}), $\xi_b(z,t)$ and
$\xi_R(z,t)$ are statically independent white Gaussian noises with
the usual properties $\langle\xi_b(z,t)\rangle=0$,
$\langle\xi_R(z,t)\rangle=0$,
$\langle\xi_b(z,t)\xi_b(z',t')\rangle=\sigma_b\delta(z-z')\delta(t-t')$,
$\langle\xi_R(z,t)\xi_R(z',t')\rangle=\sigma_R\delta(z-z')\delta(t-t')$,
where $\sigma_b$ and $\sigma_R$, are the noise intensities.
We note that the two noise sources are spatially uncorrelated, that
is at the generic point $z$ no effects is present due to random
fluctuations occurring in $z'\neq z$.\\
\subsection{Results of the stochastic model}
\begin{figure}[h]
\centering
\includegraphics[width=12cm]{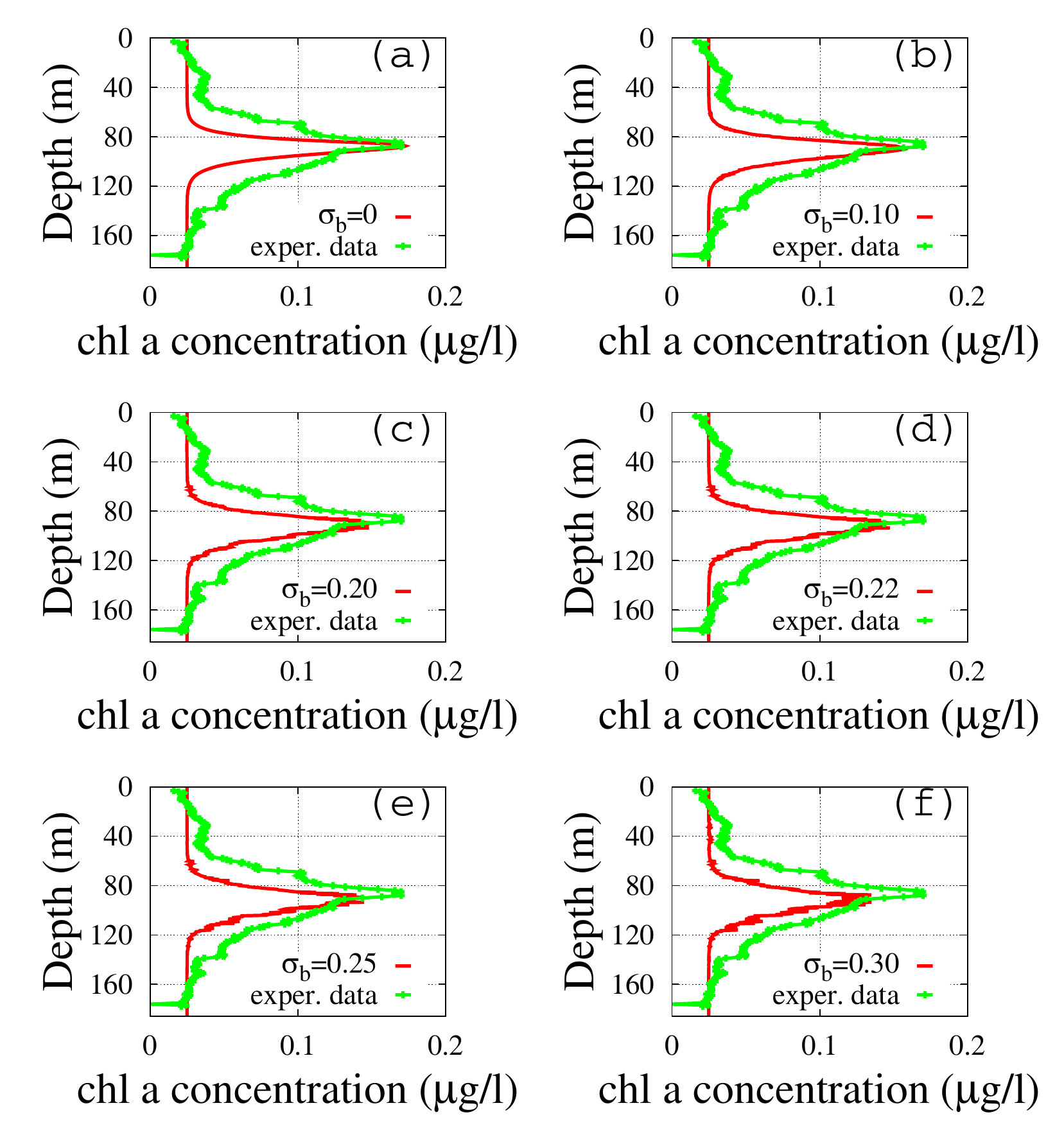}
\vspace{-0.4cm} \caption{(Color online). Average \textit{chl a}
concentration calculated (red line) for different values of
$\sigma_b$ by the stochastic model (case 1, see
Eqs.~(\ref{contornob}), (\ref{evoluzR}), (\ref{contornoR}),
(\ref{evoluzI}), (\ref{evoluzbrum})) as a function of depth. Results
are compared with \textit{chl a} distributions measured (green line)
in site L1129b. The theoretical values were obtained averaging over
$1000$ numerical realizations. The values of the parameters are
those shown in Table~\ref{table1}. The noise intensities are: (a)
$\sigma_b=0$ (deterministic case), (b) $\sigma_b=0.10$, (c)
$\sigma_b=0.20$, (d) $\sigma_b=0.22$, (e) $\sigma_b=0.25$ and (f)
$\sigma_b=0.30$.}\label{fig:site1_b_noisy}
\end{figure}

\begin{figure}[h]
\centering
\includegraphics[width=12cm]{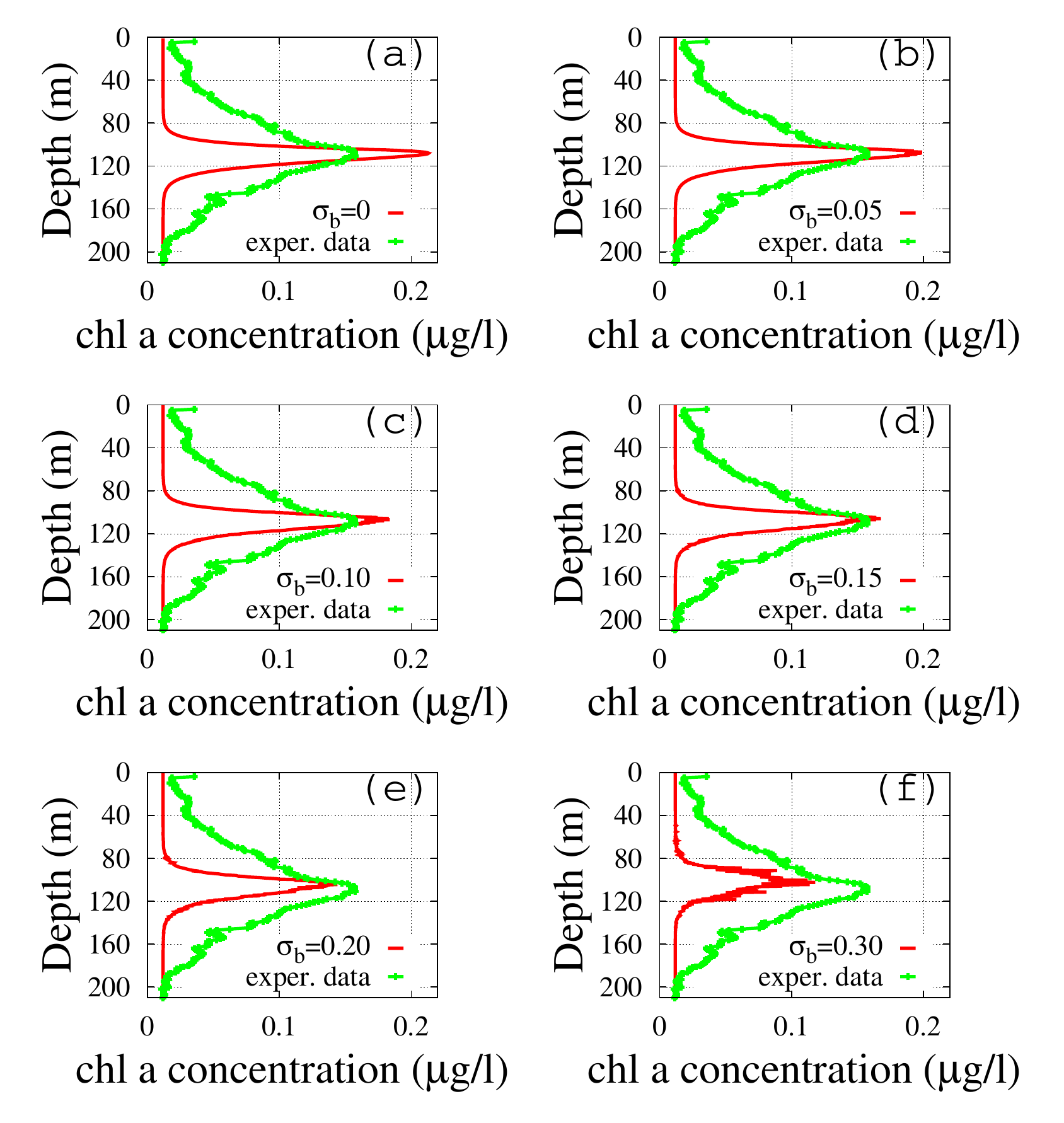}
\caption{(Color online). Average \emph{chl a} concentration
calculated (red line) for different values of $\sigma_b$ by the
stochastic model (case 1, see Eqs.~(\ref{contornob}),
(\ref{evoluzR}), (\ref{contornoR}), (\ref{evoluzI}),
(\ref{evoluzbrum})) as a function of depth. Results are compared
with \textit{chl a} distributions measured (green line) in site
L1105. The theoretical values were obtained averaging over $1000$
numerical realizations. The values of the parameters are those shown
in Table~\ref{table1}. The noise intensities are: (a) $\sigma_b=0$
(deterministic case), (b) $\sigma_b=0.05$, (c) $\sigma_b=0.10$, (d)
$\sigma_b=0.15$, (e) $\sigma_b=0.20$ and (f)
$\sigma_b=0.30$.}\label{fig:site3_b_noisy} \vspace{-0.5cm}
\end{figure}

In this paragraph we solve numerically, within the Ito scheme, the
equations of the stochastic model for different values of the noise
intensities, obtaining the distributions of the picophytoplankton
concentration as an average over $1000$ realizations. We recall that
the ecosystem is characterized by non-linear interactions among its
parts. Because of this feature the response of the system to
external solicitations is also non-linear. Therefore, one can not
expect that the presence of a symmetric noise with zero mean, i.e.
Gaussian noise used in the model, produces in average the same
effect as a deterministic dynamics~\citep{Giu09}. On the other side,
the use of a random function, i.e. noise source, to simulate the
spatio-temporal behaviour of the system, makes the single
realization unpredictable and unique, and therefore
non-representative of the real dynamics. As a consequence, one
possible choice to describe correctly the time evolution of the
system is to calculate the average of several realizations. This
procedure, indeed, allows to take into account different
"trajectories" obtained by the integration of the stochastic
equations, without focusing on a specific realization~\citep{Spa04}.
According to the discussion of Paragraph~\ref{res_det_mod}, we
calculated the solutions for a maximum simulation time $t_{max} =
4\cdot 10^{4}~h$. In Figs.~\ref{fig:site1_b_noisy}
and~\ref{fig:site3_b_noisy} we show the results for case 1.
Here we note that, in both sites, for higher noise intensities the
peaks of the two average \emph{chl a} distributions show: (i) a
decrease of their magnitude; (ii) a small displacement along the
water column. For suitable values of the noise intensity the peaks
of the average \emph{chl a} distributions obtained from the model
match very well the experimental data. We note also that the two
DCMs are located at 90 m (site L1129b) and 106 m (site L1105) of
depth (in Figs.~\ref{fig:site1_b_noisy}d and
\ref{fig:site3_b_noisy}d compare theoretical (red line) and
experimental (green line) profiles).
\begin{table*}[h]
\centering
\includegraphics[width=5cm]{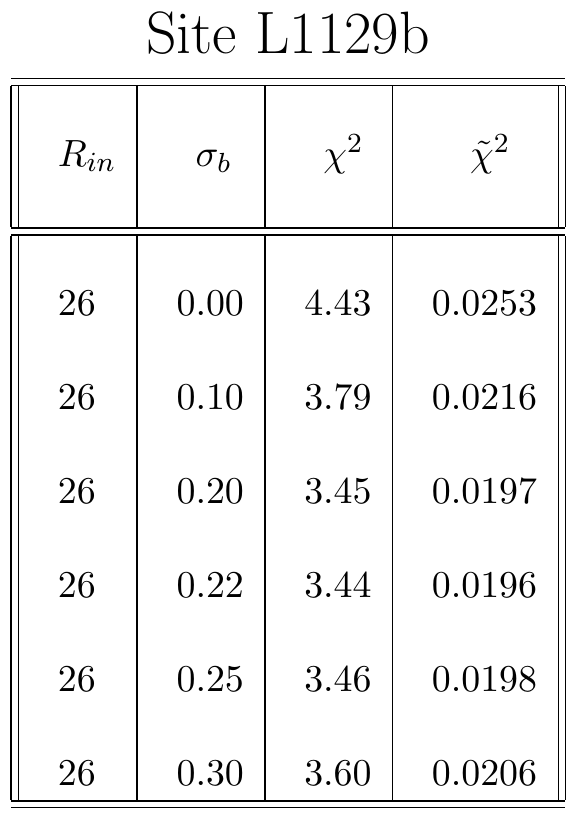}\hspace{0.5cm}
\includegraphics[width=5.15cm]{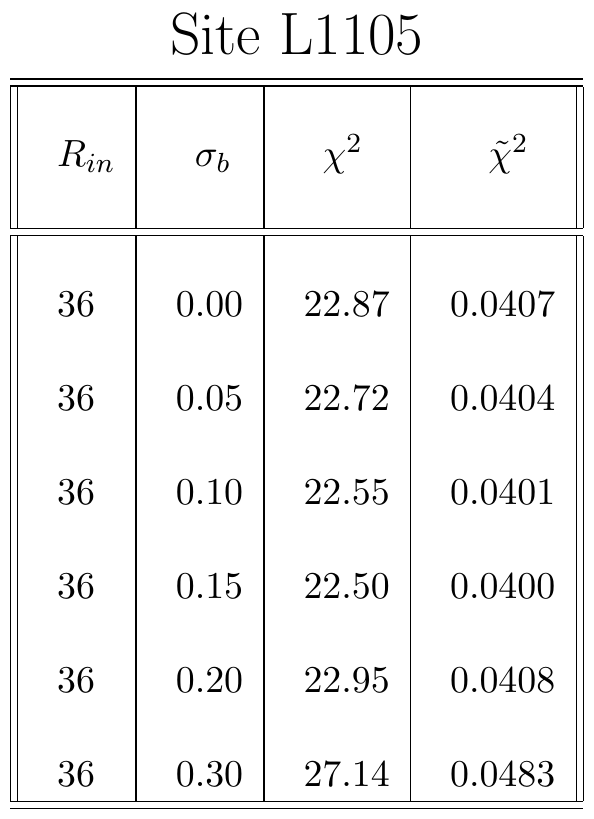}
\caption{Results of $\chi^2$, reduced chi-square ($\tilde{\chi}^2$)
goodness-of-fit test for site L1129b (left panel) and site L1105
(right panel) for different values of $\sigma_b$ (stochastic
dynamics - case 1). The number of samples along the water column is
n = 176 for site L1129b and n = 563 for site L1105.}\label{table2}
\end{table*}
\begin{figure}[h]
\centering
\includegraphics[width=4.0cm,height=3.8cm]{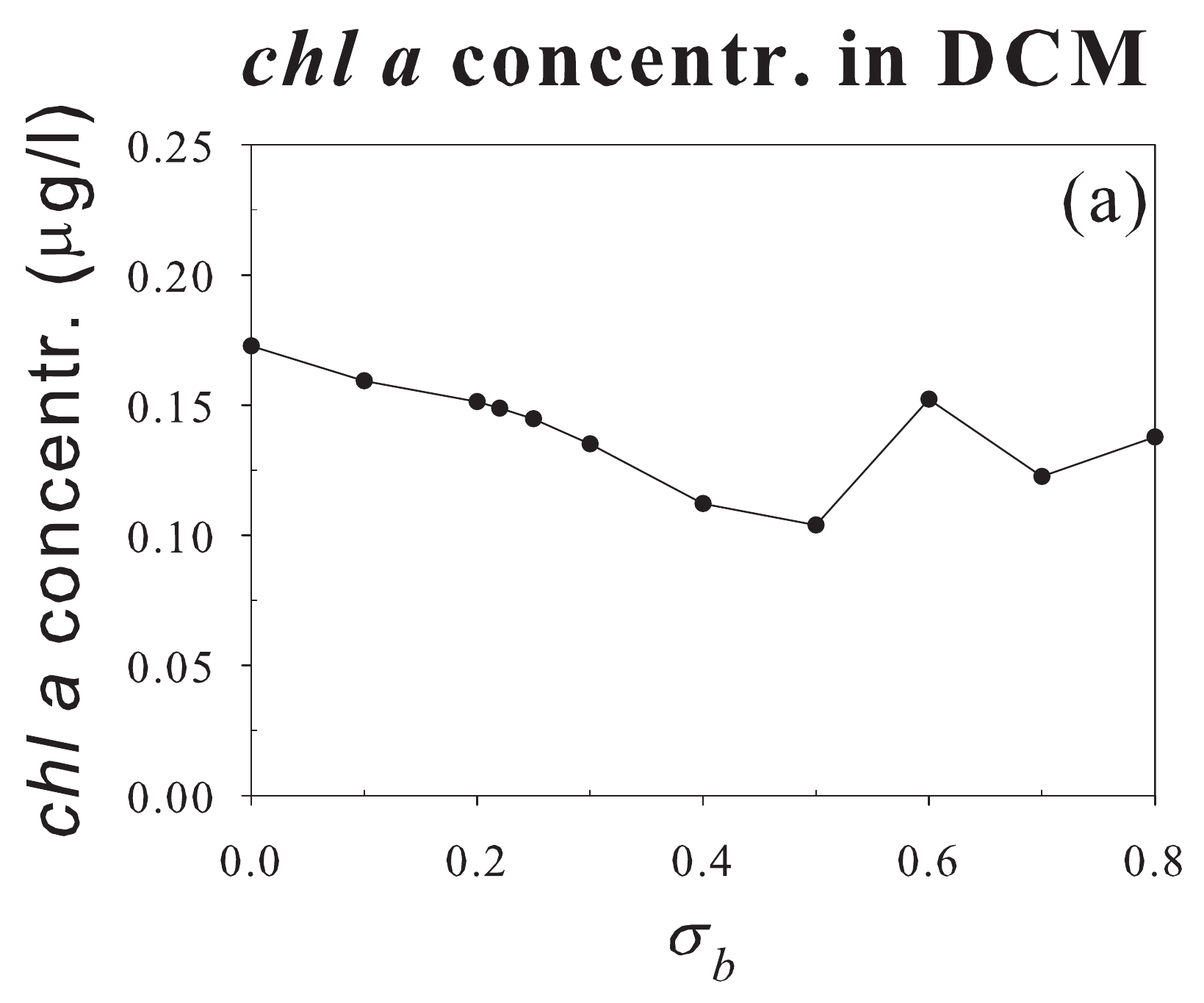}\hspace{0.1cm}
\includegraphics[width=4.0cm,height=3.8cm]{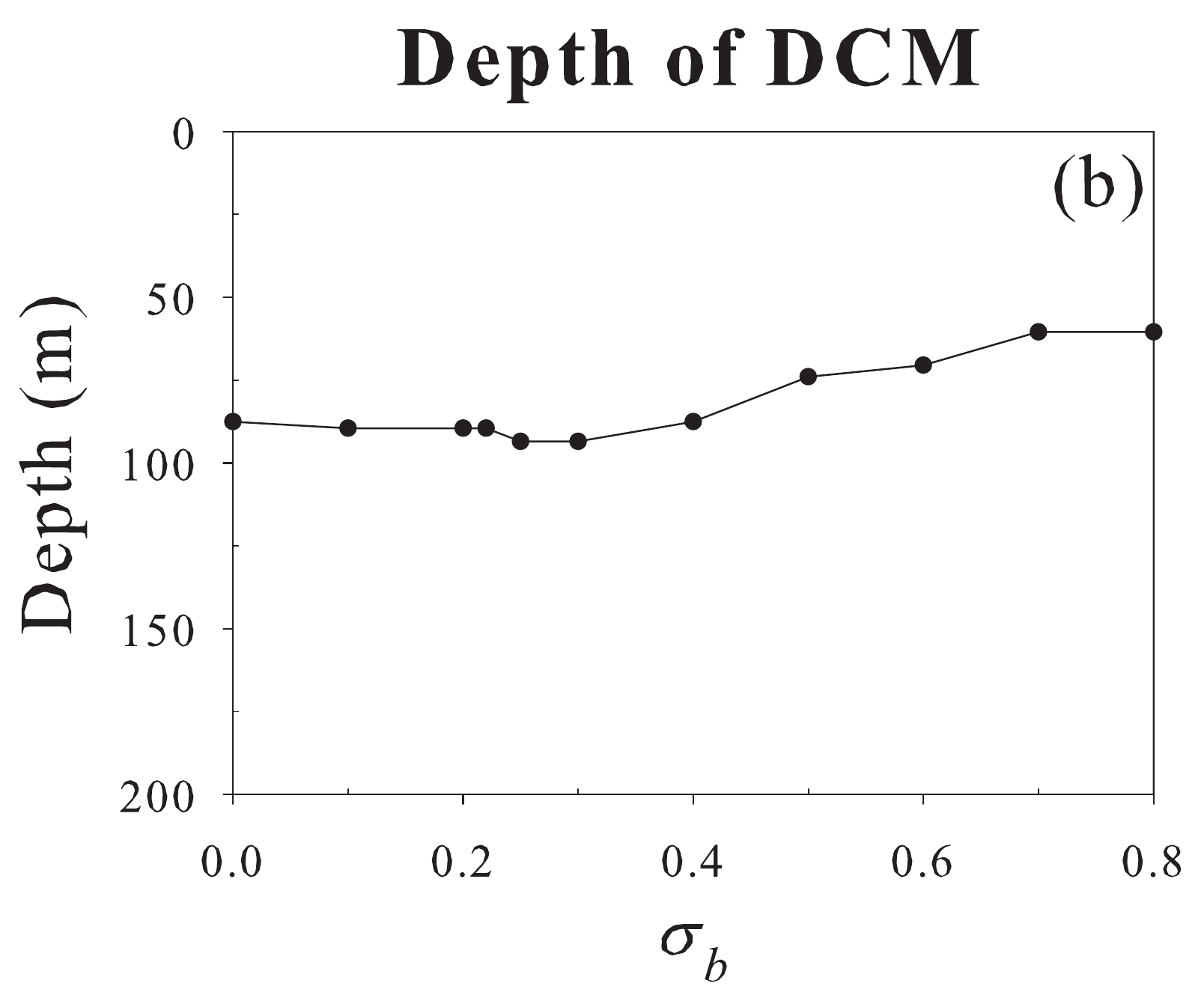}\hspace{0.1cm}
\includegraphics[width=4.0cm,height=3.8cm]{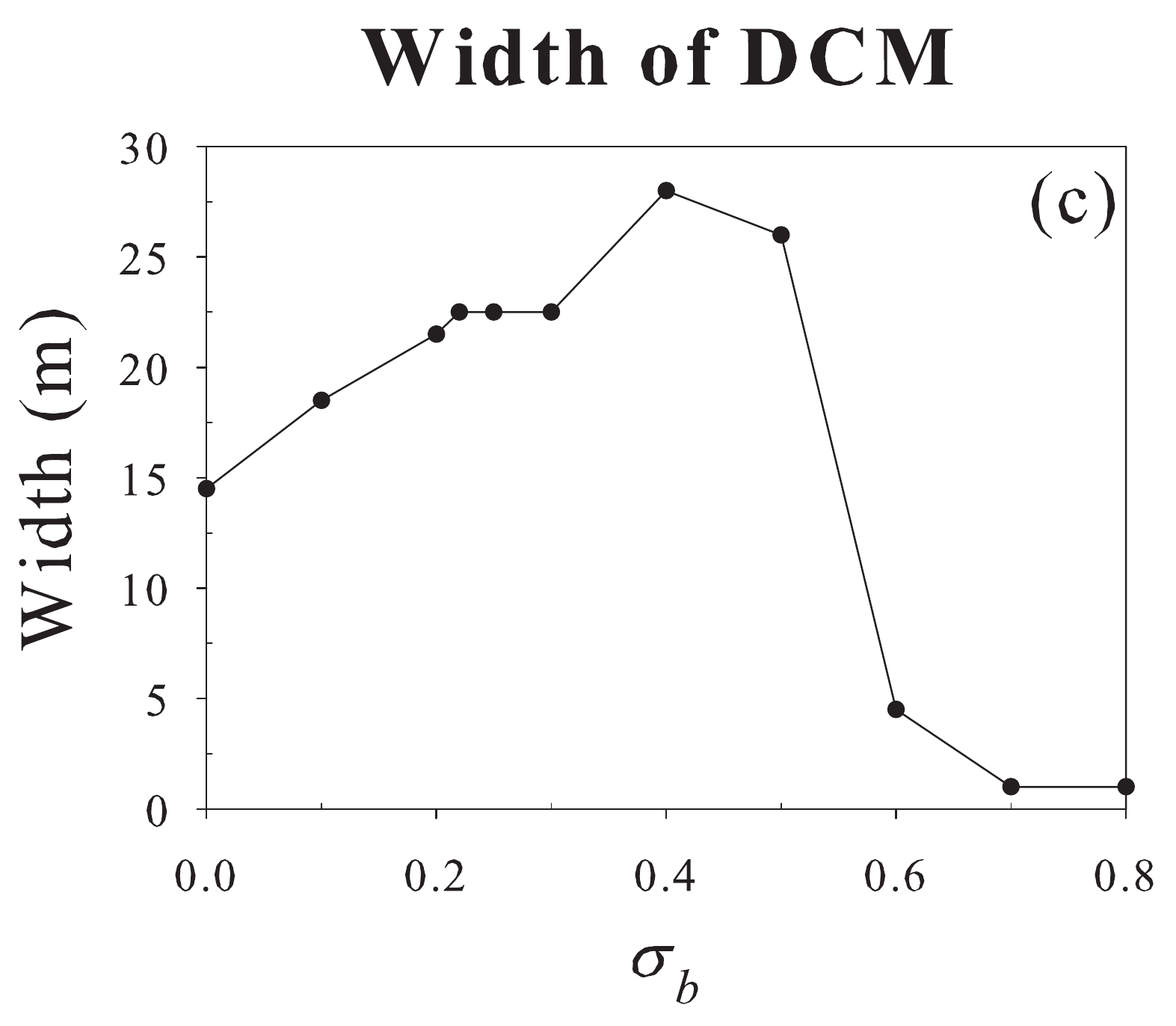}\\
\includegraphics[width=4.1cm,height=3.8cm]{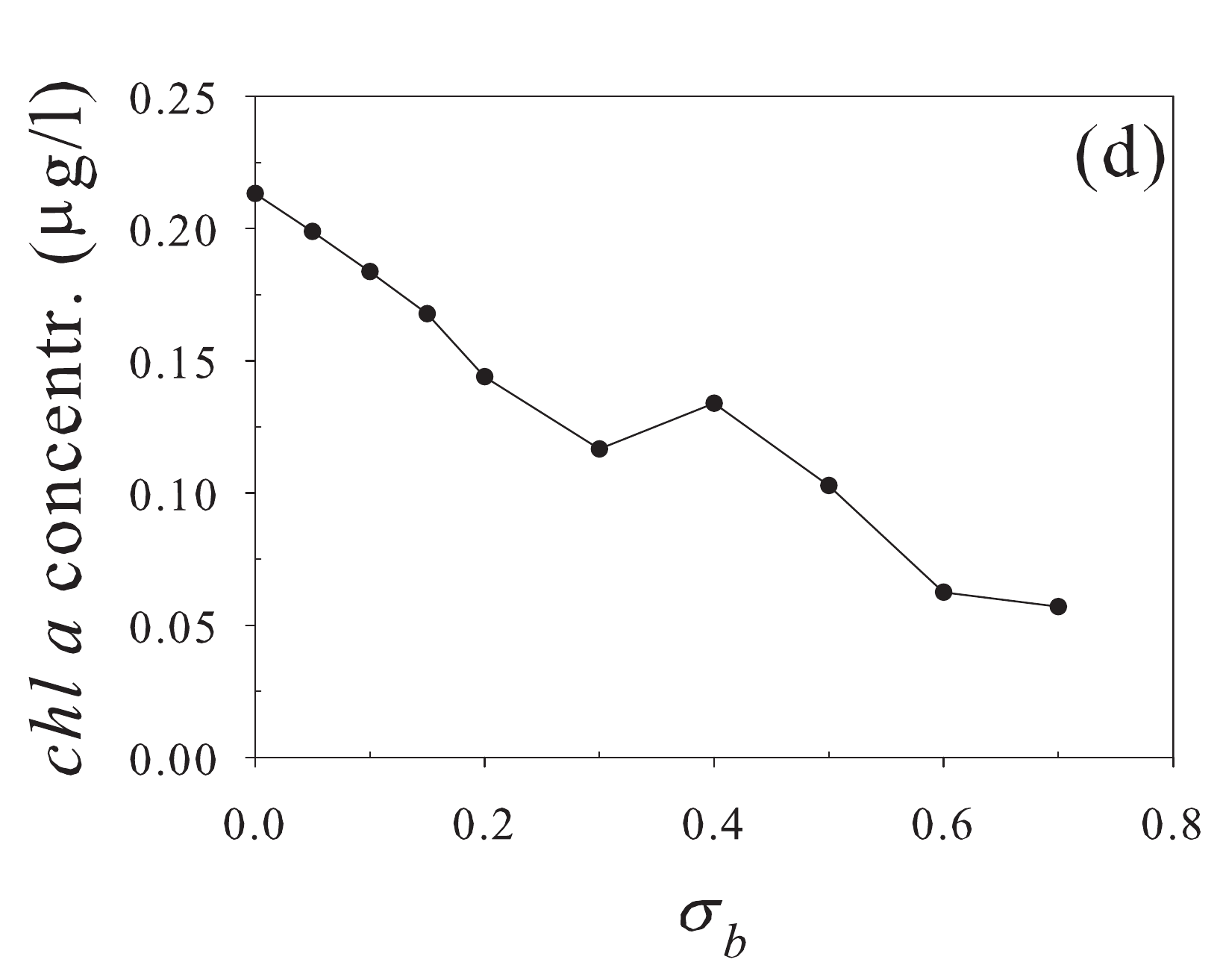}\hspace{0.1cm}
\includegraphics[width=4.1cm,height=3.8cm]{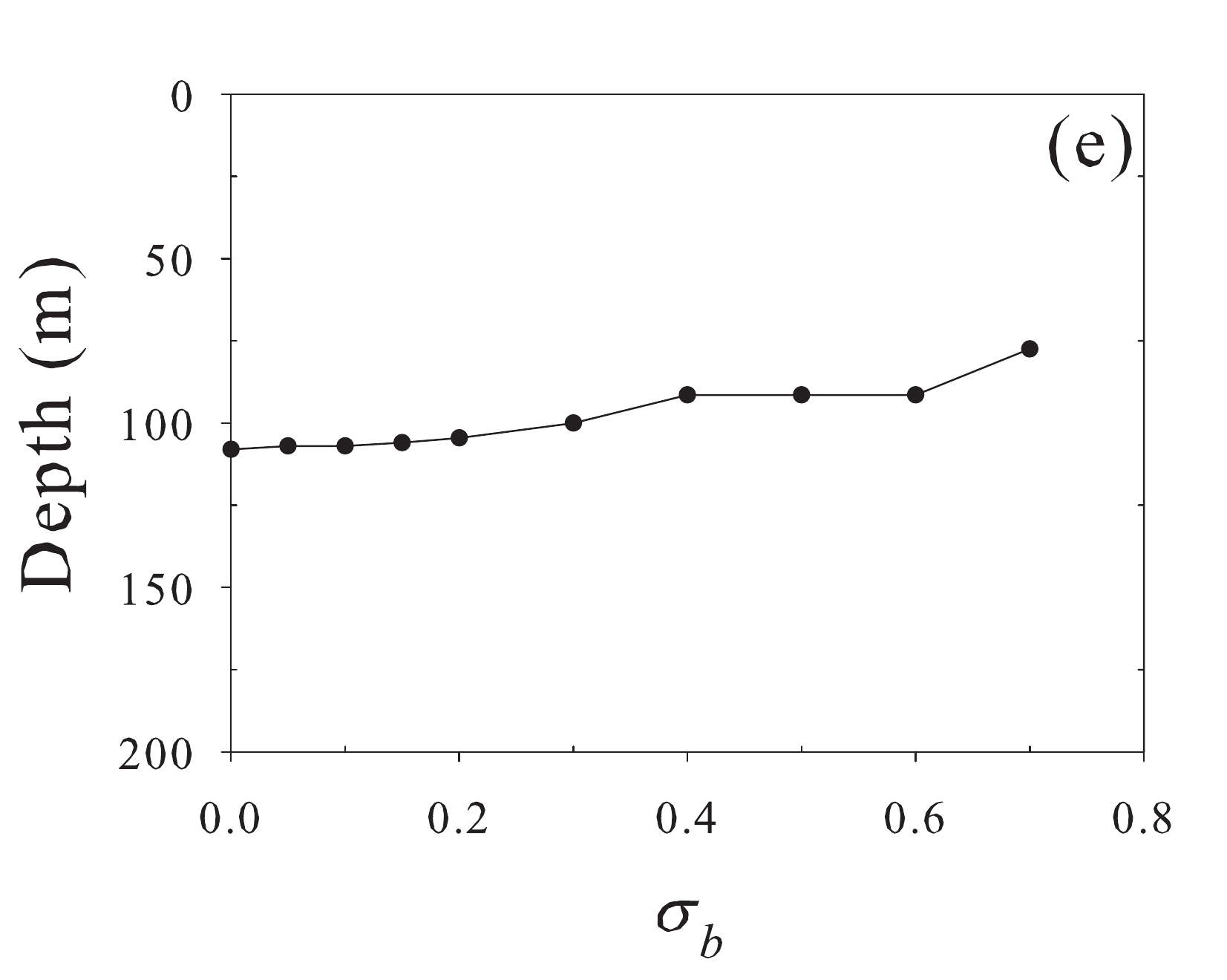}\hspace{0.1cm}
\includegraphics[width=4.1cm,height=3.8cm]{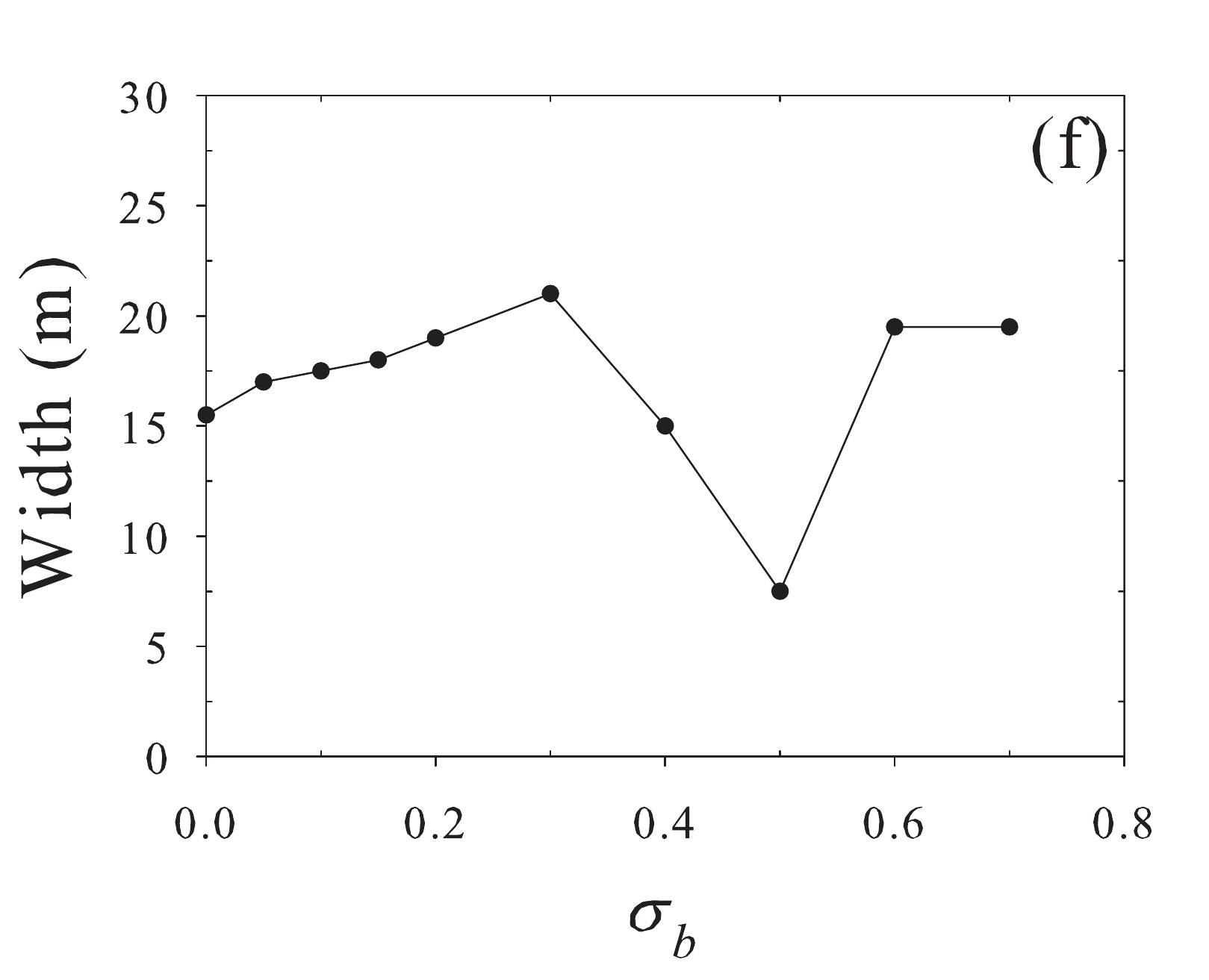}
\caption{Magnitude, depth, and width of the DCM as a function of
$\sigma_b$ obtained from the model for site L1129b (panels a, b, c)
and site L1105 (panels d, e, f).}\label{DCM_vs_sigmab}
\end{figure}
\begin{figure}[htbp]
\centering
\includegraphics[width=12cm]{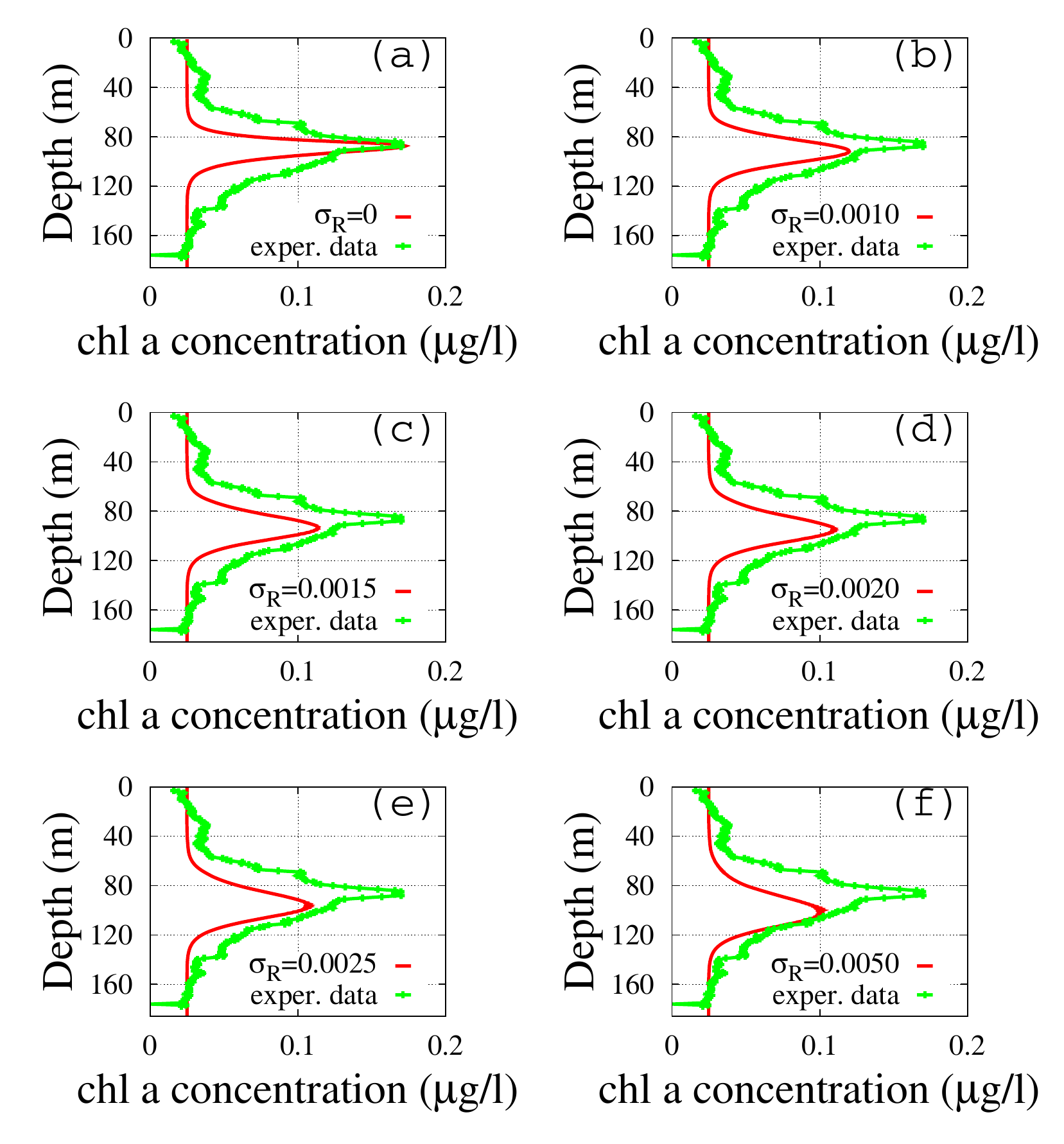}
\caption{(Color online). Average \textit{chl a} concentration
calculated (red line) for different values of $\sigma_R$ by the
stochastic model (case 2, see Eqs.~(\ref{evoluzb}),
(\ref{contornob}), (\ref{contornoR}), (\ref{evoluzI}),
(\ref{evoluzRrum})) as a function of depth. Results are compared
with \textit{chl a} distributions measured (green line) in site
L1129b. The theoretical values were obtained averaging over $1000$
numerical realizations. The values of the parameters are those shown
in Table~\ref{table1}. The noise intensities are: (a) $\sigma_R=0$
(deterministic case), (b) $\sigma_R=0.0010$, (c) $\sigma_R=0.0015$,
(d) $\sigma_R=0.0020$, (e) $\sigma_R=0.0025$ and (f)
$\sigma_R=0.0050$.}\label{fig:rumRmulti1}
\end{figure}
\begin{figure}[h]
\centering
\includegraphics[width=12cm]{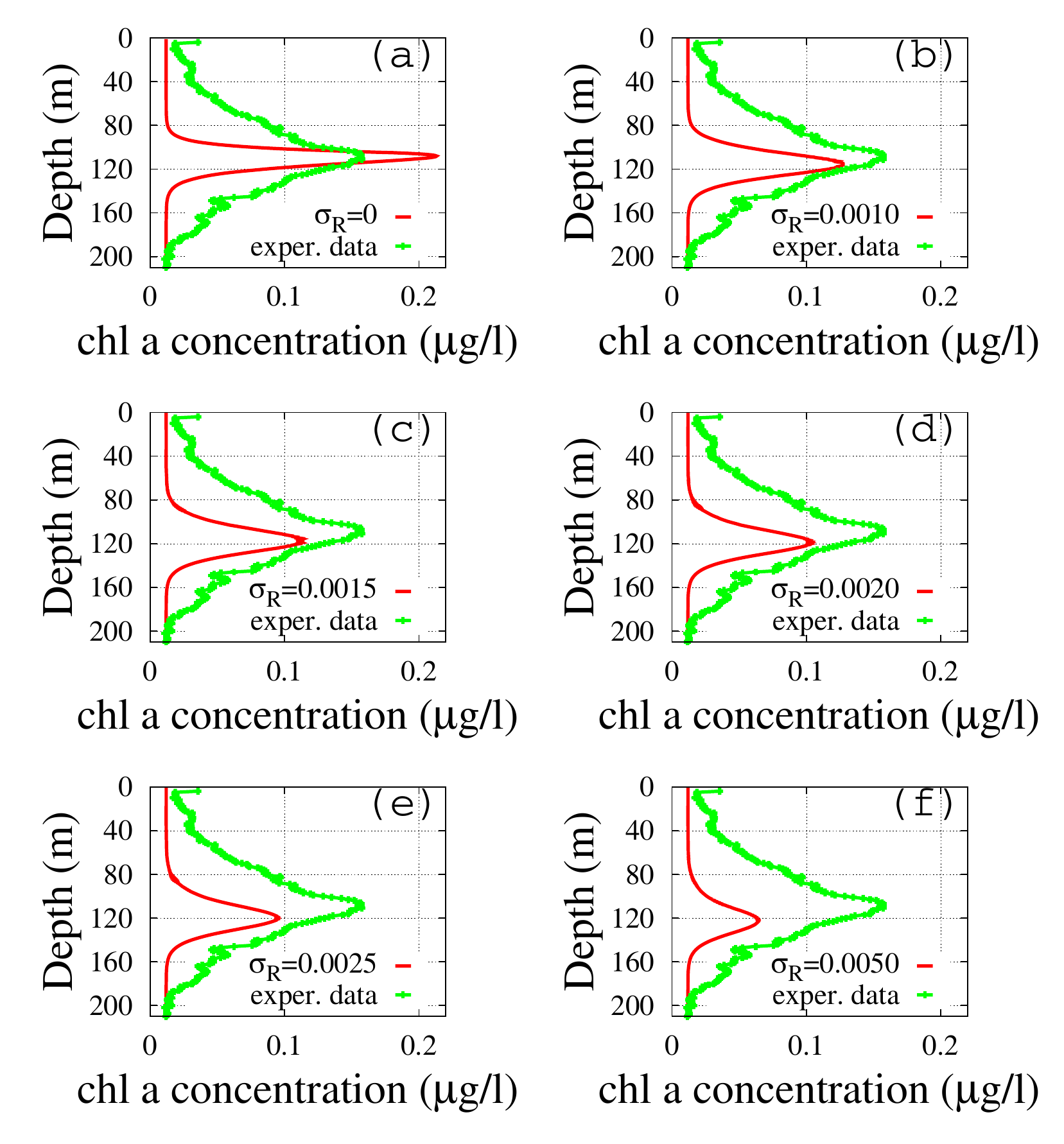}
\caption{(Color online). Average \textit{chl a} concentration
calculated (red line) for different values of $\sigma_R$ by the
stochastic model (case 2, see Eqs.~(\ref{evoluzb}),
(\ref{contornob}), (\ref{contornoR}), (\ref{evoluzI}),
(\ref{evoluzRrum})) as a function of depth. Results are compared
with \textit{chl a} distributions measured (green line) in site
L1105. The theoretical values were obtained averaging over $1000$
numerical realizations. The values of the parameters are those shown
in Table~\ref{table1}. The noise intensities are: (a) $\sigma_R=0$
(deterministic case), (b) $\sigma_R=0.0010$, (c) $\sigma_R=0.0015$,
(d) $\sigma_R=0.0020$, (e) $\sigma_R=0.0025$ and (f)
$\sigma_R=0.0050$.}\label{fig:rumRmulti3}
\end{figure}
\begin{table*}[htbp]
%[htbp]
\centering
\includegraphics[width=5cm]{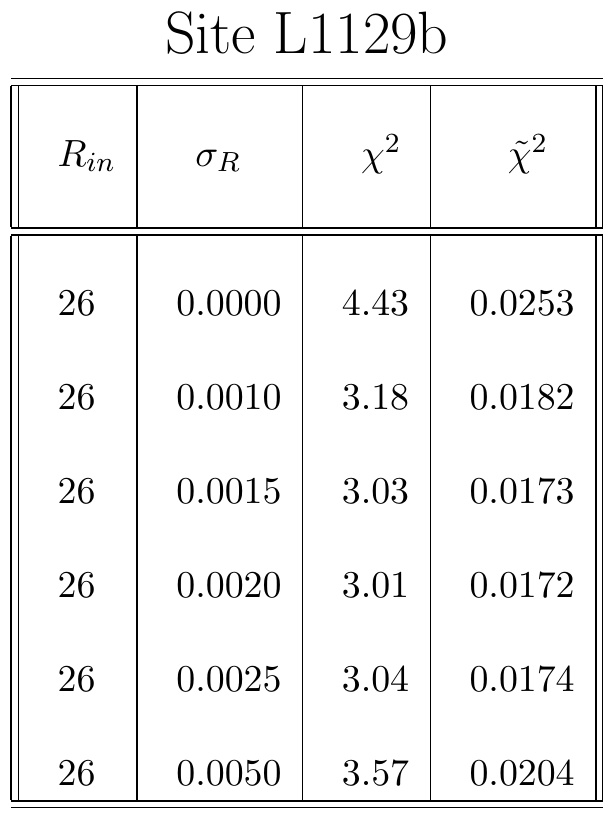}\hspace{0.5cm}
\includegraphics[width=5.23cm]{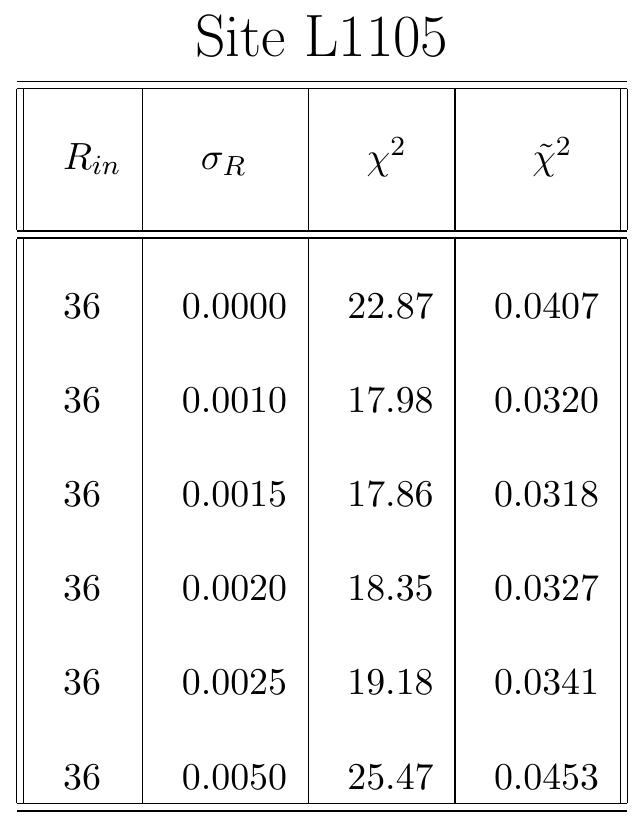}
\caption{Results of $\chi^2$, reduced chi-square ($\tilde{\chi}^2$)
goodness-of-fit test for site L1129b (left panel) and site L1105
(right panel) for different values of $\sigma_R$ (stochastic
dynamics - case 2). The number of samples along the water column is
n = 176 for site L1129b and n = 563 for site L1105.}\label{table3}
\end{table*}
To better understand the dependence of the biomass concentration on
the random fluctuations of the nutrient, according to the procedure
followed for case 1, we study for both sites the behaviour of the
depth, width, and magnitude of the DCM as a function of $\sigma_R$.
\begin{figure}[h]
\centering
\includegraphics[width=4.0cm,height=3.8cm]{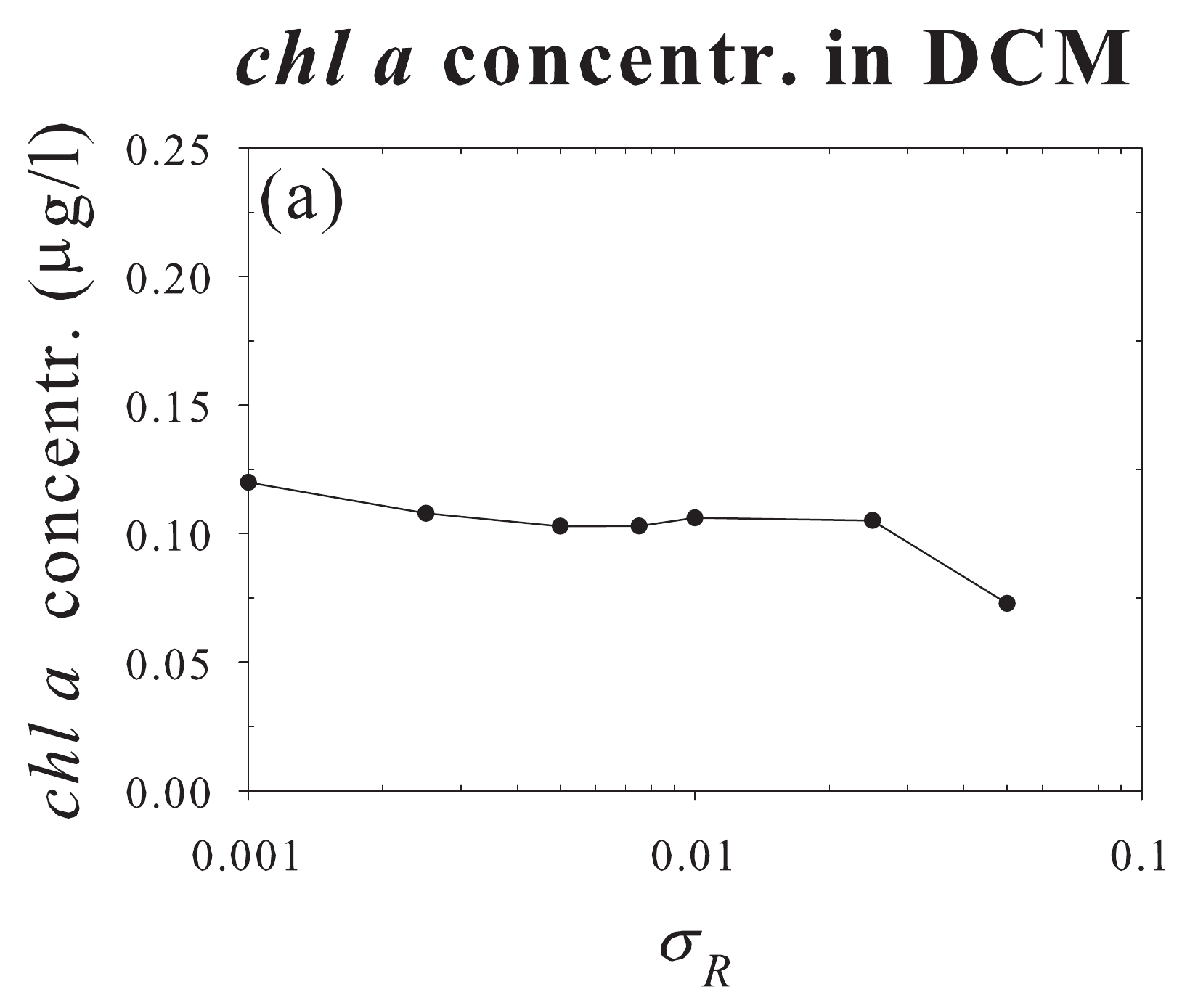}\hspace{0.1cm}
\includegraphics[width=4.0cm,height=3.8cm]{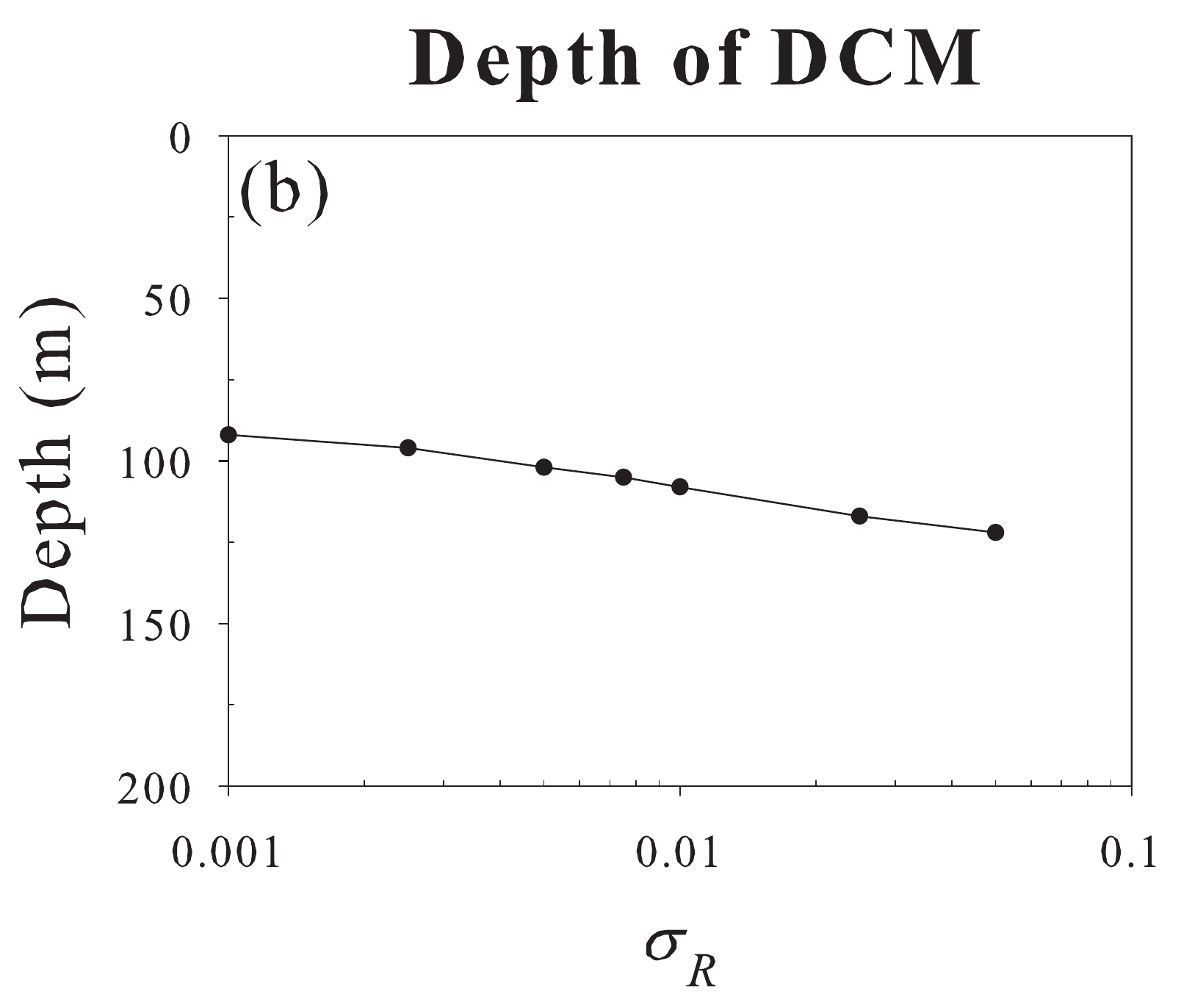}\hspace{0.1cm}
\includegraphics[width=4.0cm,height=3.8cm]{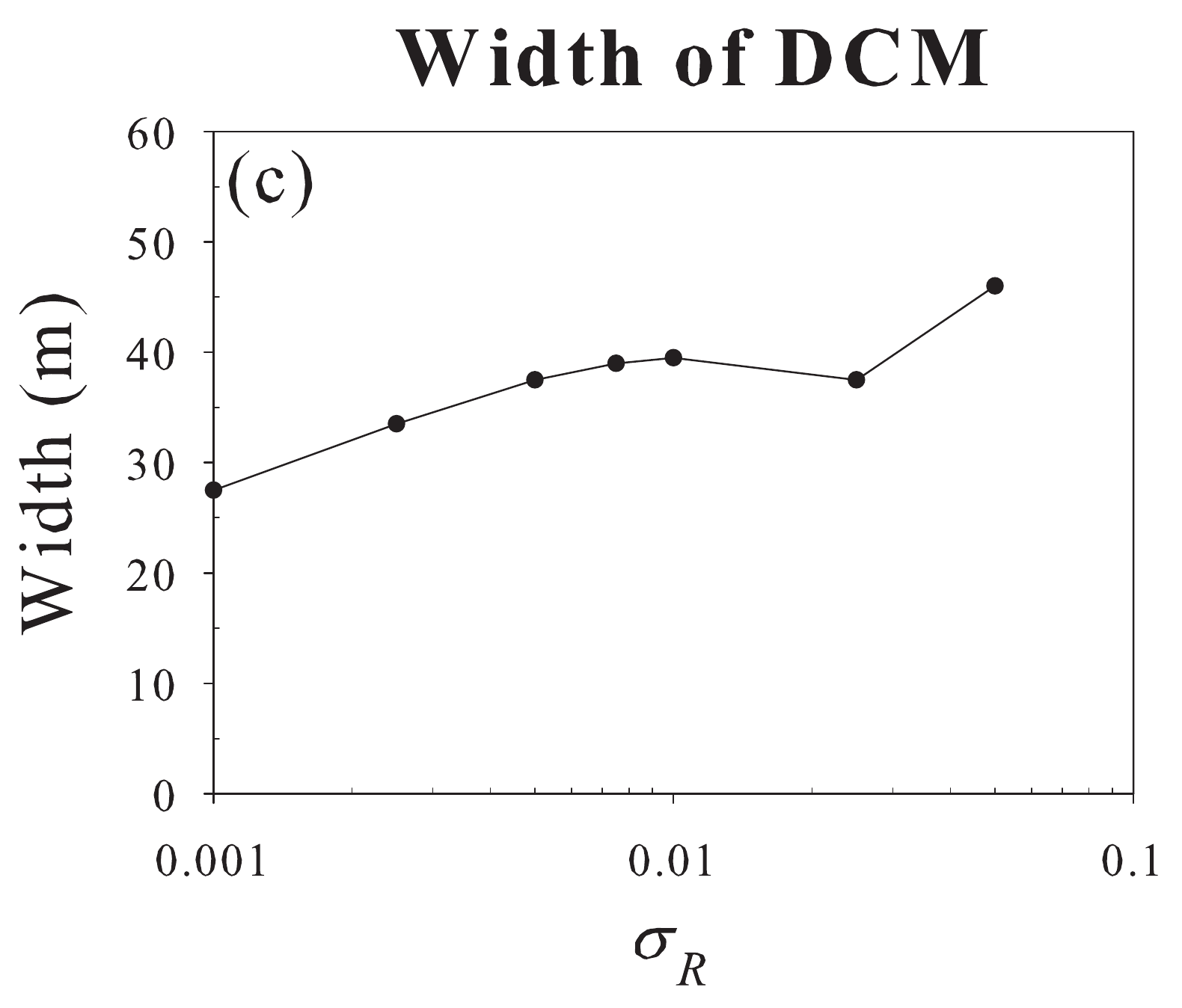}\\
\includegraphics[width=4.0cm,height=3.8cm]{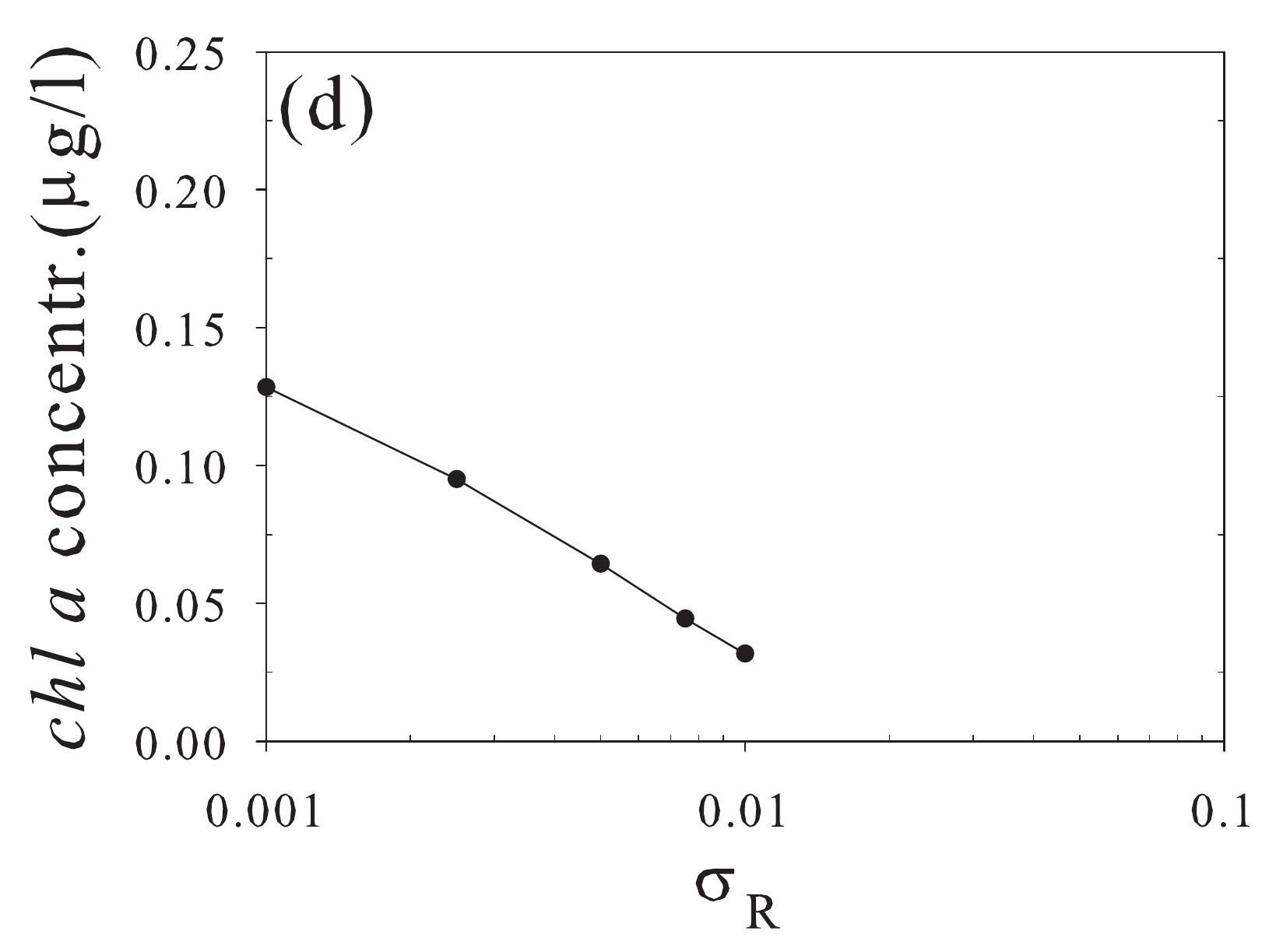}\hspace{0.1cm}
\includegraphics[width=4.1cm,height=3.8cm]{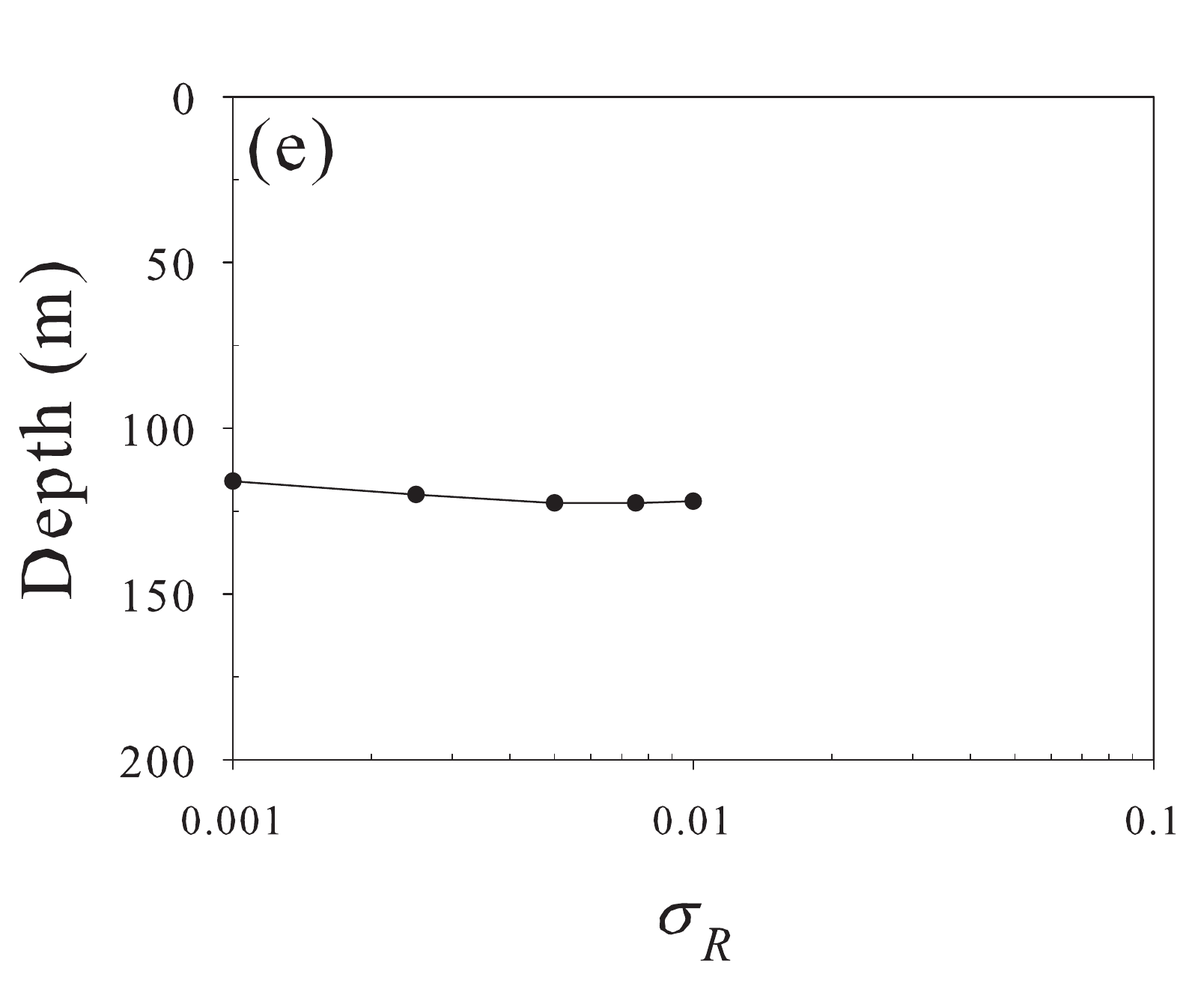}\hspace{0.1cm}
\includegraphics[width=4.1cm,height=3.8cm]{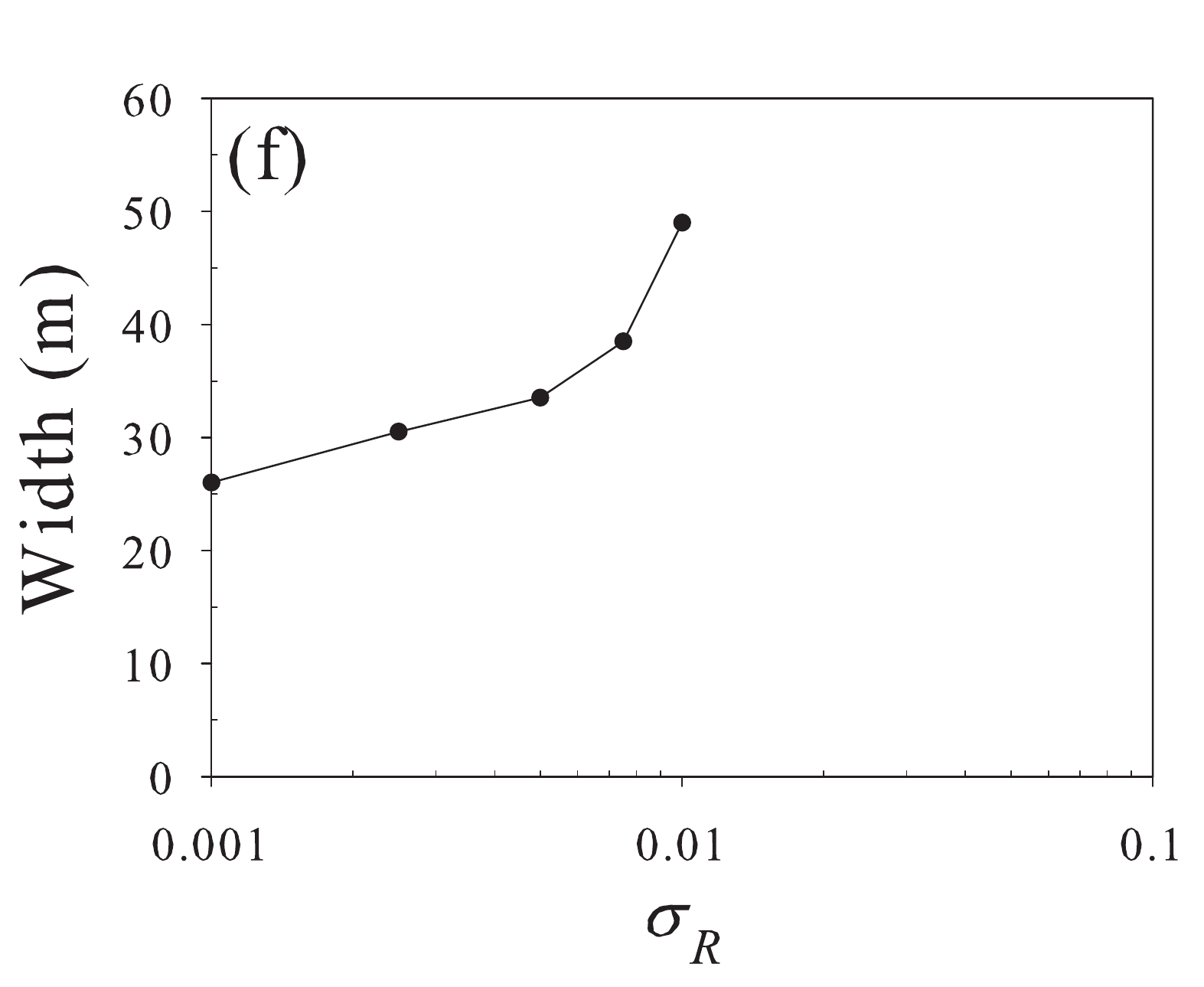}
\caption{Magnitude, depth, and width of the DCM as a function of
$\sigma_R$ obtained from the model for site L1129b (panels a, b, c)
and site L1105 (panels d, e, f).}\label{DCM_vs_sigmaR}
\end{figure}
A quantitative comparison of each theoretical \emph{chl a}
distribution (red line) with the corresponding experimental one
(green line) was carried out by performing $\chi^2$ goodness-of-fit
test. The results are shown in Tables~\ref{table2}, where
$\tilde{\chi}^2$ indicates the reduced chi-square. Results of the
$\chi^2$ test show that the smallest difference between theoretical
and experimental \emph{chl a} distributions is obtained for
$\sigma_b = 0.22$ in site L1129b and $\sigma_b = 0.15$ in site
L1105. We also note that the depths of the
DCMs are almost the same as in the deterministic case.\\
In order to better analyze this aspect, we study for both sites the
behaviour of the magnitude, depth, and width of the DCM as a
function of $\sigma_b$. The results, shown in
Fig.~\ref{DCM_vs_sigmab}, indicate that the depth of the DCM is
almost constant for $\sigma_b\leq 0.4$, increasing for higher values
of the noise intensity (see panels b, e of
Fig.~\ref{DCM_vs_sigmab}). Conversely, the width of DCM is
characterized by a non-monotonic behaviour for increasing noise
intensities. In particular, we note that the width of the DCM
exhibits a maximum in both sites (for $\sigma_b\leq 0.4$ in site
L1129b and $\sigma_b\leq 0.3$ in site L1105). For higher noise
intensities the width tends to zero for site L1129b, while a minimum
is present for site L1105 at $\sigma_b\leq 0.5$. However, for
$\sigma_b> 0.4$, the values of the DCM width are less significant,
since the chl a concentration along the water column and in
particular in the DCM decrease strongly, as can be checked in panels
a, d. In particular, random fluctuations, cause the reduction of
biomass concentration and its displacement along the water column,
determining the extinction of the picophytoplankton in the presence
of higher intensities of noise. In this condition a clear
determination of the DCM becomes more difficult. As a consequence,
the values of depth and width for the DCM are less reliable. This
analysis shows that the stationary conditions of the system depends
strongly on the environmental fluctuations, which play a critical
role in determining the best life conditions for the
picophytoplankton species.\\
\indent We complete the analysis of the stochastic dynamics,
considering the noise source which affects directly the nutrient
concentration (case 2). By numerically solving the corresponding
equations of motion (see
Eqs.~(\ref{evoluzb}),(\ref{contornob}),(\ref{contornoR}),(\ref{evoluzI}),(\ref{evoluzRrum}))
and averaging over $1000$ realizations, we obtain the average
\emph{chl a} distributions shown in Figs.~\ref{fig:rumRmulti1}
and~\ref{fig:rumRmulti3}. The results show that also for low noise
intensities ($\sigma_R$ between $0.001$ and $0.005$), a decrease and
a deeper localization of the DCMs are present. The shape of the
\emph{chl a} peaks exhibits, for both sites, a better agreement with
the corresponding experimental DCMs respect to the deterministic
case. In particular, for site L1129b the best value of the $\chi^2$
test is obtained for $\sigma_R=0.0020$, while for site L1105 the
best fitting results for $\sigma_R=0.0015$ (see Table~\ref{table3}).
We note that in site L1129b the best agreement between experimental
and numerical distributions is obtained, both in case 1 and case 2,
for values of the noise intensity, $\sigma_b$ and $\sigma_R$, higher
than those of site L1105. This can be explained by the fact that in
site L1129b the DCM is localized at a depth shallower than in site
L1105 (88m vs. 111 m), causing the environmental variables to be
subject to more intense random fluctuations due to the closer sea
surface. As a consequence, the \emph{chl a} peak in site L1129b
($88$ m) is more strongly affected by the environmental noise than
in site L1105 ($111$ m). The results, shown in
Fig.~\ref{DCM_vs_sigmaR}, indicate that the depth of the DCM
slightly increases in both sites as a function of the noise
intensity (see panels b, e of Fig.~\ref{DCM_vs_sigmaR}). We note
also that a decrease of the chl a concentration is observed in the
DCMs of the two sites. This decrease is more rapid in site L1105
(panel d), where a chl a concentration $\sim 0.025$ is reached for
$\sigma_R\sim 0.01$. Analogously we observe an increase, faster in
site L1105, of the width of the DCM. The spread of DCM and reduction
of its magnitude are strictly connected with each other. In fact,
the decrease of chl a concentration determines a flattening of the
DCM with a consequent increase of its width. In conclusion the
results shown in Fig.~\ref{DCM_vs_sigmaR} indicate that the
phytoplankton biomass tends to disappear for $\sigma_R \sim 0.01$, a
value lower than those used in case 1, where no extinction occurs up
to $\sigma_b\sim 0.7$ (see panels a, d of Fig.~\ref{DCM_vs_sigmab}).
This indicates that the stability of the nutrient concentration is a
critical factor in the dynamics of the ecosystem. Indeed, random
fluctuations of the nutrient concentration can produce dramatic
effects such as the collapse of phytoplankton biomass
considered in our model.\\
\indent The previous analysis indicates that our model is able to
reproduce the phytoplankton distributions observed in real data,
without the model taking into account explicitly the environmental
variables such as salinity and temperature. However, we observe
that, in case 2, the spatio-temporal dynamics of nutrients has been
modeled by introducing noise sources which can be interpreted as the
effect of random fluctuations of environmental variables, among
which salinity and
temperature.\\
\section{Discussion and Conclusions}
\indent In this work we presented a stochastic model, devised
starting from previous deterministic models~\citep{Lit01,Hui06}, to
study the spatio-temporal dynamics of the phytoplankton biomass
along water column in two different sites of Sicily Channel. In our
study, for fixed \emph{v}, we chose values of the vertical turbulent
diffusivity $D_{b}$ which determine the absence of intrinsic
oscillations of the phytoplankton concentration, maintaining the
system far from the chaos. In oligotrophic waters, typical of
Mediterranean Sea, where the surface mixed layer is depleted of
nutrients, subsurface maxima of chlorophyll concentration and
phytoplankton biomass are often found. Such deep chlorophyll maxima
are permanent features in large parts of the tropical and
subtropical oceans~\citep{Ven73,Cul82,Man96,Lon98,Let04}.
Furthermore, seasonal DCMs commonly develop in temperate
regions~\citep{Ven93,Lon98} and even in the polar
oceans~\citep{Hol04}, when nutrients are depleted in the surface
layer with the onset of the summer season. Here we extend recent
phytoplankton models~\citep{Kla01,Hui02,Fen03,Hod04,Hui04} to show
that the phytoplankton distributions, due to random changes, can
exhibit fluctuations.\\
\indent Our work consists in the analysis and subsequent modelling,
based on stochastic equations, of data from Sicily Channel, where
the waters are prevalently oligotrophic, the climatic conditions are
those typical of a temperate region, and the DCMs show stable
features for given conditions of light and food resources. For
values of depth ranging from $60$ to $110$ meters the presence of a
deep chlorophyll maximum indicates the existence of favourable life
conditions for the phytoplankton and results in a good agreement
with other experimental works, where higher biomass concentration
and greater diversity are observed between $60$ and $90$ meters. At
the depths considered in this work the light intensity is strongly
reduced respect to the surface value (1\% of the surface irradiance
at 75 m). However, the low light intensity did not appear to limit
the diversification of the phytoplankton
community~\citep{Bru08,Dim08}. In fact, at depths ranging from $60$
to $90$ meters a greater bio-diversity is observed. This can be
explained considering that, at these values of depth, the high
concentration of nutrients determines the most favourable life
conditions for many species of phytoplankton~\citep{Rey88}.
Differences in the composition of phytoplankton between the surface
and the DCMs are evident mainly for the smaller size class (less
than $3$ $\mu m$), which exhibits greater bio-diversity at depths
between $60$ and $90$ meters. This could be due to the fact that
different species of phytoplankton exhibit different responses to
the limiting conditions. We recall that in the marine sites analyzed
in this work the incident light intensity is characterized by high
values ($I_{in}>1300$ $\mu mol$ photon $m^{-2}s^{-1}$). Therefore,
close to the surface the low nutrient concentration represents a
limiting condition for all the phytoplankton species, so that the
biomass concentration increases with depth. However, for larger
values of depth the light intensity becomes a main limiting factor
for some species, such as Synechococcus, which show a low degree of
adaptability to smaller values of light
intensity~\citep{Moo95,Bru08}. This causes Prochlorococcus and
picoeukaryotes, which show a high degree of genetic plasticity
~\citep{Bib03} and tolerate lower light
intensities~\citep{Moo95,Moo98,Dim07}, to exhibit a dominance in the
deep chlorophyll maximum~\citep{Bru07}.\\
\indent In our model, the values of the biological parameters are
those of the picoeukaryotes and the environmental parameters are set
at values typical of the oligotrophic waters during the warm period.
These values allow to obtain \emph{chl a} distributions along the
water column in a good agreement with the experimental data and
provide limiting conditions typical of the south part of
Mediterranean Sea during the summer. Changes in the phytoplankton
composition, both qualitatively and quantitatively, are related to
the different depths considered, with light intensity and nutrient
availability being the most important factors. Picophytoplankton
demonstrated greater ability for photoacclimation than nano- and
micro-phytoplankton~\citep{Bru03,Bru06,Bru07,Dim07,Bru08,Dim08}. In
fact, a higher contribution of picoeukaryotes to the phytoplankton
biomass is observed, specifically pelagophytes and prymnesiophytes,
which were also found to thrive elsewhere in cyclonic
eddies~\citep{Ola93,Vai03}. This ability was also observed in
culture~\citep{Dim07,Dim08,Dim09}.\\
\indent On the basis of our theoretical findings we can conclude
that the position of the deep chlorophyll maximum depends on the
parameter values used in the model. We used values of the buoyancy
velocity \emph{v} and vertical turbulent diffusivity $D_{b}$, for
which no oscillations occur. In this work we used the condition
$D_b=D_R=0.5 \thinspace cm^{2}/s$, corresponding to poorly mixed
waters along the whole water column, which causes the phytoplankton
peak to have a width of few meters, as observed in the experimental
data. Moreover, we also considered in our model the presence of an
upper mixed layer, above the thermocline, characterized by a higher
value of the diffusion coefficients ($D_b=D_R=50 \thinspace
cm^{2}/s$), keeping $D_b=D_R=0.5$ $cm^{2}/s$ for greater
depth~\citep{Rya10}. The results (here not shown) did not evidence
any variations in the picophytoplankton distributions respect to the
case of uniform diffusion coefficients ($D_b=D_R=0.5$ $cm^{2}/s$)
along the whole water column. This can be explained noting that in
the ecosystem considered here the mixed layer, due to the depth of
the thermocline, is not enough thick to influence the DCMs of the
chlorophyll distributions.\\
In our ecosystem the position and stability of the chlorophyll
maximum, obtained from the model, depend not only on the vertical
turbulent diffusivity, but also on the nutrient concentration at the
bottom $R_{in}$ and the maximum specific growth rate $r$. We also
note that the values of $R_{in}$ used in our model are compatible
with the nutrient concentrations measured along the
water column in several sites of the Mediterranean Sea~\citep{Rib03,Bru06,Bru07}.\\
Our numerical results were calculated by setting the maximum
specific growth rate $r$ at a value consistent with experimental
observations. Specifically, this value has been chosen so that the
net per capita growth rate $g(z,t)$, used in the model, is in a good
agreement with those experimentally observed for
the picoeukaryotes~\citep{Jac01,Tim05,Dim09}.\\
\indent We recall that the estimations of the \textit{chl a} content
per picoeukaryote cell are highly variable, depending on the depth
and water properties (oligotrophic or eutrophic) examined. Moreover
these estimations reflect the taxonomic, ecological and
physiological diversity and the plasticity highlighted in previous
studies ~\citep{Moo00,Moo01,Not05,Tim05}. In our model we took into
account this aspect. In particular, after obtaining the numerical
results for the phytoplankton concentration expressed in number of
cells/$m^3$, we used the experimental findings given in
Ref.~\citep{Bru07} to convert the numerical results into \emph{chl
a} concentration expressed in $\mu g/l$. Specifically, because of
the peculiarities of our model, suitable to describe the dynamics of
the picoeucaryotes, we used the conversion curves typical of these
species and compared the results with the experimental \emph{chl a}
concentrations sampled in two different sites of the Mediterranean
Sea (Channel of Sicily). From the comparison we found that the
values of \emph{chl a} concentration obtained numerically are in a
good agreement not only with our data but also with those measured
by Brunet et al.~\citep{Bru07}. In addition, we note that our
numerical results for the picoeukaryote concentration expressed in
number of cells/$m^3$ match the corresponding experimental data
reported in Refs.~\citep{Bru06,Bru07}.\\
\indent More precisely, as a first step we used a deterministic
model, consisting of an auxiliary equation for the light intensity
and two differential equations, one for the dynamics of the
phytoplankton biomass, the other for the dynamics of the nutrients.
The numerical results showed a good qualitative agreement with the
real data, even if discrepancies were observed between the
characteristics of the \emph{chl a} concentration profiles provided
by the model and those obtained from the real data.\\
\indent To improve the agreement between numerical and experimental
distributions, we modeled the random fluctuations of the
environmental variables, by adding a term of multiplicative Gaussian
noise in the differential equation for the phytoplankton biomass.
The results obtained indicate that the presence of random
fluctuations, acting directly on the phytoplankton biomass,
determines \emph{chl a} stationary distributions more similar to the
experimental ones. In particular, we found that both the position
and magnitude of the DCMs agree very well with the experimental
findings. Afterwards, we modified the deterministic model
considering the role of a noise source which influences directly the
dynamics of the nutrients, by adding a term of multiplicative
Gaussian noise in the differential equation for the nutrients. In
this case we observed for suitable noise intensities (much lower
than those used in the equation for the phytoplankton biomass) a
further improvement of the numerical distributions of \emph{chl a}
concentration respect to the experimental ones. In addition, we
found that higher noise intensities (comparable with those used in
the equation for the phytoplankton biomass), cause a rapid
extinction of the phytoplankton community. The results obtained
indicate that the proposed stochastic model is able to reproduce
patterns of real phytoplankton distributions when aquatic ecosystems
with poorly mixed waters are considered.

\section*{Acknowledgments}

Authors acknowledge the financial support by ESF Scientific
Programme "Exploring the Physics of Small Devices (EPSD)"
coordinated by Prof. Christian Van den Broeck. This work received
also the financial support of MIUR and \emph{Geogrid Project}
managed by Prof. Goffredo La Loggia.

%\bibliographystyle{apsrev}

%\bibliography{phyto}  % list here all the bibliographies that
                        % you need.

\end{document}